\documentclass[modern]{aastex631}

\newcommand{\ecss}{mW~m$^{-2}$~sr$^{-1}$}
\newcommand{\lam}{$\lambda$}
\newcommand{\kms}{km~s$^{-1}$}
\newcommand{\as}{$^{\prime\prime}$}
\renewcommand{\ion}[2]{#1\,\textsc{#2}}
\newcommand{\hinode}{Hinode}

\submitjournal{ApJ}

\graphicspath{{./}{figures/}}

\begin{document}

\title{The Temperature and Density of a Solar Flare Kernel Measured from Extreme Ultraviolet Lines of O\,IV}

\author[0000-0001-9034-2925]{Peter R. Young}
\affiliation{NASA Goddard Space Flight Center, 
Greenbelt, MD 20771, USA}
\affiliation{Northumbria University, Newcastle upon Tyne, NE1 8ST, UK}

\begin{abstract}
Previously-unexplored diagnostics of \ion{O}{iv} in the extreme ultraviolet region 260--280~\AA\ are used to derive a temperature and density for a solar flare kernel observed on 2012 March 9 with the Extreme ultraviolet Imaging Spectrometer on the Hinode satellite.  Seven lines from the $2s2p^2$--$2s2p3s$ transition array between 271.99 and 272.31~\AA\ are both temperature and density sensitive relative to the line at 279.93~\AA. The temperature, $T$, is constrained with the \lam268.02/\lam279.93 ratio, giving a value of $\log\,(T/\mathrm{K})=5.10 \pm 0.03$. The ratio \lam272.13/\lam279.93 then yields an electron number density, $N_\mathrm{e}$, of $\log\,(N_\mathrm{e}/\mathrm{cm}^{-3}) = 12.55$ with a lower limit of 11.91, and an upper limit of 14.40. The \ion{O}{iv} emitting volume is estimated to be 0.4\as\ (300\,km) across. Additional \ion{O}{iv} lines at 196, 207 and 260~\AA\ are consistent with the derived temperature and density but have larger uncertainties from the radiometric calibration and blending. Density diagnostics of \ion{O}{v} and \ion{Mg}{vii} from the same spectrum are consistent with a constant pressure of $10^{17.0}$\,K\,cm$^{-3}$ through the transition region. The temperature derived from \ion{O}{iv} supports recent results that \ion{O}{iv} is formed around 0.15~dex lower at high densities compared to standard ``zero-density" ionization balance calculations.
\end{abstract}

\section{Introduction} \label{sect.intro}

The standard model for solar flares \citep{2017LRSP...14....2B} invokes energy release in the corona by magnetic reconnection to produce energetic particles and thermal conduction fronts that propagate to the solar surface along magnetic field lines. Energy is dissipated in the chromosphere giving rise to bright flare ribbons in H$\alpha$, for example. Emission along the ribbons is not uniform and there appear bright ``kernels" likely due to localized enhanced heating. The chromospheric plasma can be heated to temperatures $>$ 10\,MK, and so the ribbons and kernels can be seen in a wide range of ultraviolet emission lines. 

Models of energy transfer and heating along 1D flare loops such as RADYN \citep{2015ApJ...809..104A} are able to predict the evolution of temperature and density in a flare loop at high cadence. From observations we seek to provide constraints on these parameters using line ratio diagnostics. To apply line ratio diagnostics, it is necessary to consider optically-thin lines, which limits diagnostics to ions formed above around 50,000 K.

Generally we expect flare loops to evolve under constant pressure, and so the highest densities will be at the lowest temperatures. Previous work has found transition region densities of $10^{11}$\,cm$^{-3}$ to $10^{13}$\,cm$^{-3}$ \citep{1987ApJ...320..913W,2016A&A...594A..64P,2018ApJ...857....5Y}. The number of line ratio diagnostics sensitive in this  range is very limited, but the intercombination lines of \ion{O}{iv} and \ion{S}{iv} near 1400 A have been applied to data from the currently-operating Interface Region Imaging Spectrograph \citep[IRIS:][]{2014SoPh..289.2733D} mission and the earlier Skylab \citep{1987ApJ...320..913W} and Solar Maximum missions \citep{1987ApJ...312..943H}. The \ion{O}{iv} diagnostics are sensitive up to about $10^{12.5}$\,cm$^{-3}$, and the \ion{S}{iv} diagnostics are sensitive to $10^{13}$\,cm$^{-3}$. \citet{2016A&A...594A..64P} provided an example of a flare observed by IRIS that exhibited densities at the maxima of these ranges. It is possible to infer higher densities than allowed by these ions by comparing an intercombination line with a resonance line, with \ion{O}{iv} \lam1401.16 / \ion{Si}{iv} \lam1402.77 a particularly attractive ratio as the lines are nearby and close in formation temperature. It has been applied to flare data by \citet{1977ApJ...215..652F}, \citet{2016A&A...594A..64P} and \citet{2018ApJ...857....5Y} but it is subject to many more uncertainties than the single-ion ratios, as discussed in these articles.

Modern UV and EUV solar spectrometers are generally confined to relatively narrow wavelength ranges due to optical design limitations, the size of electronic detectors, and the need for high spectral resolution. There is an approximate dividing line at 400~\AA\ such that coronal lines formed at temperatures of 1~MK and above lie at shorter wavelengths and chromospheric and transition region lines at longer wavelengths \citep{2021FrASS...8...50Y}. This limits the ability of the spectrometers to make density measurements over a wide temperature range. The Extreme ultraviolet Imaging Spectrometer \citep{2007SoPh..243...19C} on Hinode falls on the coronal side of this dividing line and has a number of density diagnostics in the 0.7--6~MK region that have previously been applied to flares \citep[e.g.,][]{2008ApJ...685.1277W,2011ApJ...740...70M}. However, the EIS wavelength ranges of 170--212\,\AA\ and 246--292\,\AA\ contain a number of emission lines of \ion{O}{iv} through \ion{O}{vi} that extend the EIS temperature coverage to the mid-TR. The lines are much weaker than these ions' main resonance lines in the 500--1100~\AA\ region, but they are measurable in many circumstances and have been used for diagnostics \citep[e.g.,][]{2010ApJ...708..550M}.

In this article a surprisingly rich spectrum of \ion{O}{iv} lines in an EIS spectrum of a flare kernel is presented and used to provide constraints on temperature and density. This extends the EIS diagnostic coverage to a temperature region that is normally the preserve of longer wavelength spectrometers such as IRIS. 

Section~\ref{sect.soft} describes the availability of data and software used for this article. Section~\ref{sect.model} discusses the atomic model for \ion{O}{iv} and the available emission lines and diagnostics. The EIS flare dataset is described in Section~\ref{sect.obs} and the data processing steps are given. Section~\ref{sect.meas} explains how the \ion{O}{iv} lines are measured, and Section~\ref{sect.diag} presents the results from the diagnostic ratios. A comparison with diagnostics from other ions is given in Section~\ref{sect.comp}, and Section~\ref{sect.fchroma} discusses the results in relation to a flare model. Finally, results are summarized in Section~\ref{sect.summary}.

\section{Data and Software}\label{sect.soft}

Spectra from EIS are analyzed in the present work. EIS was launched on board the \textit{Hinode} spacecraft in 2006, and it is an imaging slit spectrometer operating in the wavelength bands 170--212~\AA\ and 246--292~\AA---referred to as the short-wavelength (SW) and long-wavelength (LW) channels, respectively. Four slit options are available, including the narrow 1\as\ and 2\as\ slits for emission line spectroscopy. Spectral resolution is 3000--4000 and spatial resolution is 3\as\--4\as\ \citep{2022SoPh..297...87Y}. EIS level-0 data are  available from the  Virtual Solar Observatory\footnote{\url{https://virtualsolar.org}.}, and are calibrated to level-1 using routines in the \textit{SolarSoft} \citep{1998SoPh..182..497F,2012ascl.soft08013F} distribution.

Images from the Atmospheric Imaging Assembly \citep[AIA:][]{2012SoPh..275...17L} are used to provide context images and identify the position of the EIS slit in relation to the rapidly developing flare ribbon. AIA obtains full-disk solar images in seven EUV channels at 12~s cadence. Cutout images were downloaded from the \href{https://jsoc.stanford.edu}{Joint Science Operations Center} and respiked as flare ribbon pixels are often incorrectly flagged as energetic particle hits by the AIA processing pipeline \citep{2013ApJ...766..127Y}. After respiking, the images were then cleaned of energetic particle hits with the routine \textsf{aia\_clean\_cutout\_sequence} \citep{2021SoPh..296..181Y}.

Most software used for data preparation and analysis for this article are available in \textit{Solarsoft}. Additional software of the author for AIA analysis are available in a GitHub repository\footnote{\url{https://github.com/pryoung/sdo-idl}.}.
Atomic data and software for computing emissivities are taken from version 10.1 of the CHIANTI atomic database \citep{2016JPhB...49g4009Y,2023ApJS..268...52D}, which is distributed through \textit{Solarsoft}.
The spectra and line fit data from this article are available on \href{https://zenodo.org}{Zenodo} at doi:\href{https://doi.org/10.5281/zenodo.6726426}{10.5281/zenodo.6726426}. 

\section{The O IV Atomic Model and Spectrum}\label{sect.model}

\begin{figure*}[t]
    \centering
    \includegraphics[width=\textwidth]{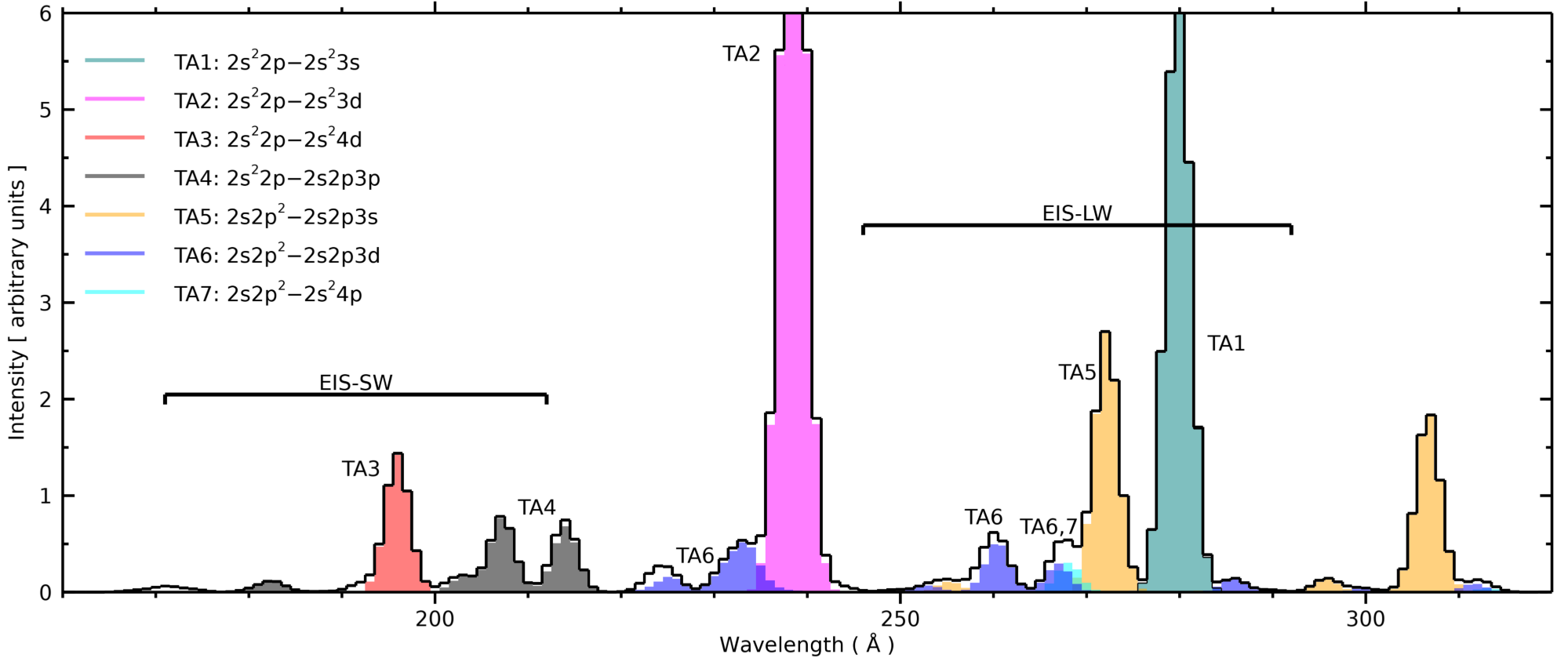}
  \caption{A CHIANTI \ion{O}{iv} spectrum computed at $\log\,(T/\mathrm{K})=5.15$ and $\log\,(N_\mathrm{e}/\mathrm{cm}^{-3})=9.36$ with line widths set to 3\,\AA. Transition arrays (TAs) for the spectral features are indicated, and the wavelength ranges of the EIS SW and LW channels are highlighted with black horizontal lines.}
  \label{fig.chianti}
\end{figure*}

\ion{O}{iv} has a rich spectrum of emission lines through the far and extreme ultraviolet. Transitions between $n=2$ configurations ($2s^22p$, $2s2p^2$) give rise to the strongest emission lines with three key groups of resonance lines at 553--556~\AA, 608--610~\AA\ and 787--791~\AA. The lines observed by EIS and discussed in the present work arise from $n=3$ and $n=4$ configurations. Figure~\ref{fig.chianti} and Table~\ref{tbl.trans} summarize these lines, and further details are given later in this Section.

Version~10.1 of CHIANTI was used for modeling the \ion{O}{iv} emission lines in this article.  Experimental energies come from the NIST database, \citet{1986ApJS...61..801S} and \citet{1997ApJ...487..962F}. 
Radiative decay rates are from \citet{2004ADNDT..86...19C} and \citet{2012A&A...547A..87L}, and electron collision strengths are  from \citet{2012A&A...547A..87L}. For the lines in the 260--280~\AA\ range, the NIST energies were derived from wavelength measurements of \citet{edlen34} for which the uncertainties  range from 6 to 11~m\AA.

\ion{O}{iv} is formed at middle transition region temperatures. The equilibrium ionization balance data provided by CHIANTI gives a peak ionization at $\log\,(T_\mathrm{iz}/\mathrm{K})=5.15$. Non-Maxwellian particle distributions, which are likely during flares, can significantly modify the ionization fraction distribution of \ion{O}{iv} \citep{2014ApJ...780L..12D}, but for the high density found here (Section~\ref{sect.diag}) such distributions would be expected to be quickly thermalized by collisions. For a specific \ion{O}{iv} transition, the temperature of formation is better represented by the temperature at which the contribution function peaks ($T_\mathrm{cf}$). 
The transitions considered in the present work have high excitation potentials, giving larger emissivities at higher temperatures. They therefore have formation temperatures of $\log\,(T_\mathrm{cf}/\mathrm{K})=5.25$ to 5.30. For a specific solar feature for which a differential emission measure (DEM) curve is defined, a third formation temperature comes from multiplying the contribution function by the DEM and temperature \citep[see, for example,][]{peter_r_young_2023_8097457}. This temperature is referred to as $T_\mathrm{dem}$ here. Since the standard quiet Sun, active region and flare DEMs distributed with CHIANTI show DEM curves that decrease with increasing temperature through the \ion{O}{iv} formation region, then $T_\mathrm{dem}$ will be lower than $T_\mathrm{cf}$. The issue of the formation temperature of \ion{O}{iv} lines is returned to in Section~\ref{sect.diag}.

EUV density diagnostics arise from population shifting from the ground level to higher excitation levels as the increasing electron density causes greater electron excitation rates. The population of an excited level increases until it reaches a quasi-Boltzmann equilibrium (QBE) with the ground level. \ion{O}{iv} has  two $^2P_{1/2,3/2}$ levels in the ground configuration and these reach QBE at $\log\,(N_\mathrm{e}/\mathrm{cm}^{-3})=4.0$, which is too low for solar atmosphere density diagnostics. The next excited levels are the  three $^4P_{1/2,3/2,5/2}$ levels in the $2s2p^3$ configuration. These come into QBE with the ground level over $\log\,(N_\mathrm{e}/\mathrm{cm}^{-3})=9$ to 12, which is ideal for the solar atmosphere. The intercombination transitions coming directly from the quartet levels are found in the 1397--1408~\AA\ wavelength range and show density sensitivity relative to each other. They
have been extensively applied in both solar and stellar spectra \citep[e.g.,][]{1987ApJ...312..943H,2002MNRAS.337..901K,2016A&A...594A..64P}.

Any quartet term at higher excitation energies will become principally excited from the $2s2p^3$ $^4P$ levels as they gain population, and hence transitions from these levels will form useful density diagnostics in relation to lines from doublet levels that are principally excited from the ground $^2P$ term. An example is the set of three $2s2p^2$ $^4P_J$ -- $2p^3$ $^4S_{3/2}$ transitions at 625~\AA, that form excellent diagnostics with the set of four $2s^22p$ $^2P_J$ -- $2s2p^2$ $^2P_{J^\prime}$ transitions near 554~\AA. These lines were observed with the Coronal Diagnostic Spectrometer \citep{1995SoPh..162..233H} and \citet{1997SoPh..175..523Y} presented an observation of an intense transition region brightening that yielded a density of $\log\,(N_\mathrm{e}/\mathrm{cm}^{-3})=11.3$.

\begin{deluxetable}{ccccD}[t]
\tablecaption{\ion{O}{iv} transitions in the EIS wavelength bands. Transition arrays (TAs) correspond to those shown on Figure~\ref{fig.chianti}.\label{tbl.trans}}
\tablehead{
  \colhead{TA} &
  \colhead{Terms} &
  \colhead{$J$--$J^\prime$} &
  \colhead{$\lambda$\tablenotemark{a} (\AA)} &
  \twocolhead{$I$\tablenotemark{b}} \\
}
\decimals
\startdata
TA3 & $^2P$--$^2D$ & 1/2--3/2 & 195.860 & 10.7 \\
 & & 3/2--5.2 & 196.006 & 19.7 \\
 & & 3/2--3/2 & 196.008 & 2.2 \\
\noalign{\smallskip}
TA4 & $^2P$--$^2D$ & 1/2--3/2 & 207.183 & 5.8 \\
 & & 3/2--5/2 & 207.239 & 10.6 \\
 & & 3/2--3/2 & 207.348 & 1.2 \\
\noalign{\smallskip}
TA6a & $^2D$--$^2F$ & 5/2--7/2 & 260.389 & 6.8 \\
 & & 3/2--5/2 & 260.556 & 4.7 \\
\noalign{\smallskip}
TA6b & $^2D$--$^2D$ & 5/2--5/2 & 266.931 & 3.6 \\
 & & 3/2--3/2 & 266.981 & 2.4 \\
\noalign{\smallskip}
TA7 & $^2D$--$^2P$ & 5/2--3/2 & \uline{268.024} & 4.2 \\
 & & 3/2--1/2 & 268.064 & 2.3 \\
\noalign{\smallskip}
TA5 & $^4P$--$^4P$ & 3/2--5/2 & 271.990 & 9.6 \\
 & & 1/2--3/2 & 272.076 & 8.4 \\
 & & 5/2--5/2 & \uline{272.127} & 22.4 \\
 & & 3/2--3/2 & 272.173 & 2.7 \\
 & & 1/2--1/2 & 272.176 & 1.6 \\
 & & 3/2--1/2 & 272.273 & 7.8 \\
 & & 5/2--3/2 & 272.310 & 9.1 \\
\noalign{\smallskip}
TA1 & $^2P$--$^2S$ & 1/2--1/2 & 279.631 & 49.9 \\
 & & 3/2--1/2 & \uline{279.933} & 100.0 \\
\noalign{\smallskip}
\enddata
\tablenotetext{a}{The lines used for density and temperature diagnostics (Figure~\ref{fig.dens}) are underlined.}
\tablenotetext{b}{Intensity computed with CHIANTI and normalized such that \lam279.93 has an intensity of 100.}
\end{deluxetable}

Figure~\ref{fig.chianti} shows a synthetic \ion{O}{iv} spectrum in the region 160--300~\AA\ derived with CHIANTI and computed at $\log\,(T/\mathrm{K})=5.15$ and $\log\,(N_\mathrm{e}/\mathrm{cm}^{-3})=9.36$. Seven transition arrays (TA) are identified corresponding to transitions between different sets of configurations. A low resolution is used for the spectrum to better differentiate the TAs.  Terms, wavelengths, and theoretical intensities for the lines in the EIS wavelength bands are given in Table~\ref{tbl.trans}. The theoretical intensities are the same as those used to derive the spectrum in Figure~\ref{fig.chianti}, but normalized so that the intensity of the line at 279.93~\AA\ is 100.

The strongest line in the displayed spectrum is from TA2, but is outside the EIS wavebands. The next strongest is from TA1 at 279.93~\AA\ that is observed by EIS, and has been used for Doppler shift measurements by \citet{2010ApJ...711...75L} and \citet{2013ApJ...766..127Y}, and for temperature and density diagnostics by \citet{2010ApJ...708..550M} and \citet{2010ApJ...711...75L}. 

\begin{figure*}[t]
    \centering
    \includegraphics[width=\textwidth]{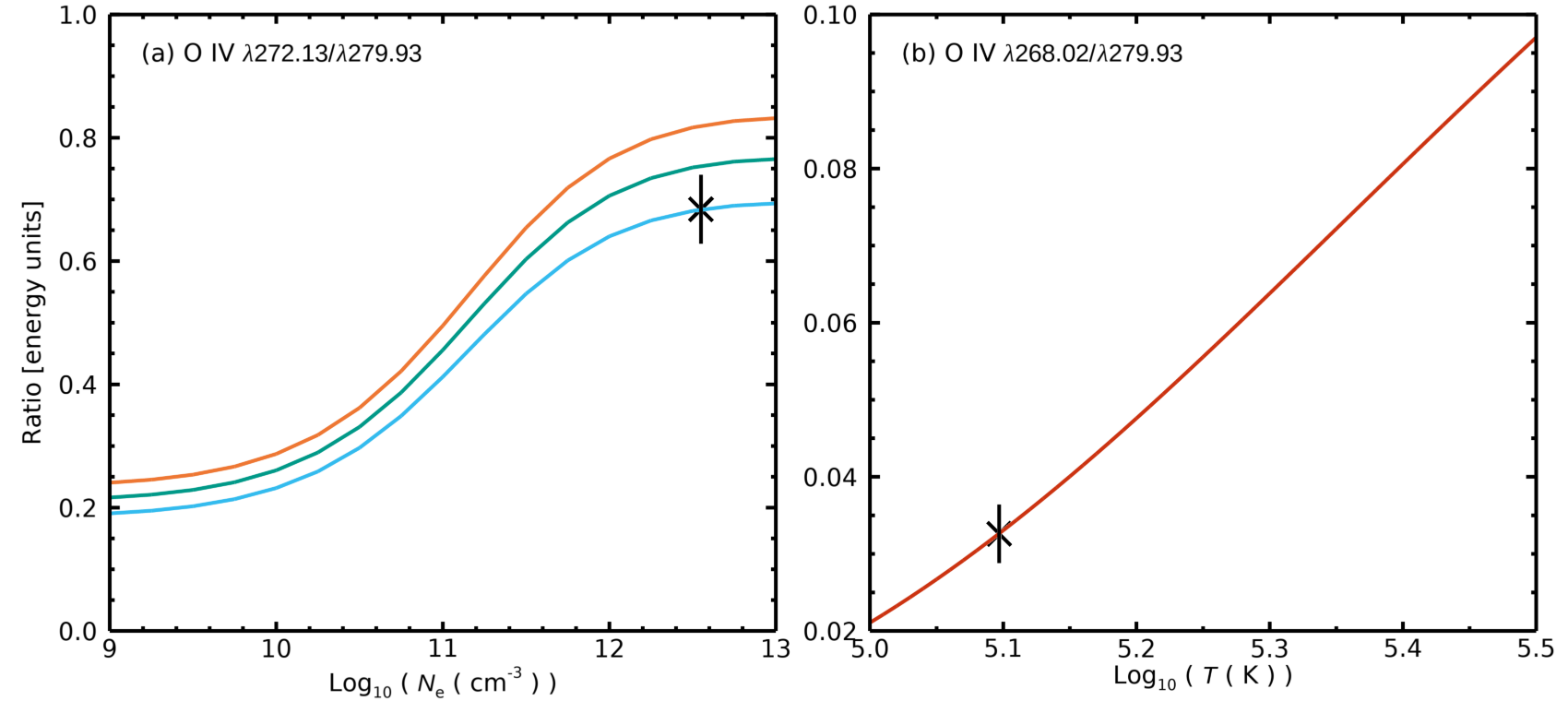}
    \caption{Theoretical \ion{O}{iv} line ratios from CHIANTI. (a) The \lam272.13/\lam279.93 ratio as a function of density, calculated for three temperatures: $\log\,(T/\mathrm{K})=5.10$ (cyan), 5.15 (teal) and 5.20 (orange). (b) The \lam268.02/\lam279.93 ratio as a function of temperature, calculated for $\log\,(N_\mathrm{e}/\mathrm{cm}^{-3})=12.0$. The crosses with error bars show the values derived from the flare kernel (Section~\ref{sect.diag}).}
    \label{fig.dens}
\end{figure*}

As discussed above, density diagnostics can be formed from ratios of quartet-quartet transitions to doublet-doublet transitions. The only quartet-quartet transitions belong to the TA5 multiplet with lines around 272~\AA, and these lines are conveniently close in wavelength to the 279.93~\AA\ line. Figure~\ref{fig.dens}(a) shows the predicted variation of \lam272.13/\lam279.93 with density obtained with CHIANTI. However, it can be seen that the ratio also has some sensitivity to temperature. To accurately derive the density it is therefore necessary to also constrain the plasma temperature, and this is possible with some of the other lines observed by EIS.

The TA6 and TA7 transitions arise from atomic levels with excitation energies around 40\,\%\ higher than that of the TA1 transitions. The Boltzmann factor in the expression for the electron excitation rate therefore means that the ratios of the TA6 and TA7 lines to the TA1 lines are strongly sensitive to temperature. Figure~\ref{fig.dens}(b) shows the \lam268.02/\lam279.93 ratio, which uses a TA7 transition. As both lines are doublet-doublet transitions, then there is negligible density sensitivity.


The combination of \lam268.02, \lam279.93 and the lines at 272~\AA\ together allow the temperature and density to be constrained. The fact that the lines are relatively close together in the spectrum helps mitigate  uncertainties in the EIS radiometric calibration (Appendix~\ref{app.calib}).

\section{Observation and Data Preparation}\label{sect.obs}

The active region with NOAA designation 11429 crossed the solar disk during 2012 March 3--14 and was one of the most intensively-studied of Solar Cycle 24. A search of abstracts in the SAO/NASA Astrophysics Data System for ``11429" yields 33 journal articles, many of which focus on the coronal mass ejections from the region \citep[e.g.,][]{2014ApJ...788L..28L,2015ApJ...809...34C,2020ApJ...901...40D}. One X-class flare occurred on March 5 and two more on March 7. Additional flares with classifications of M5 and higher occurred on March 9, 10 and 13. In the present article EIS data of the M6.3 class flare on 9 March are analyzed. The evolution of the  X-ray 1--8~\AA\ light curve obtained by the GOES-15 spacecraft was complex \citep{2013ApJ...767...55D}, with an initial rise at 03:24~UT, followed by three maxima at 03:27~UT, 03:34~UT and 03:53~UT. 

\begin{figure*}[t]
    \centering
    \includegraphics[width=\textwidth]{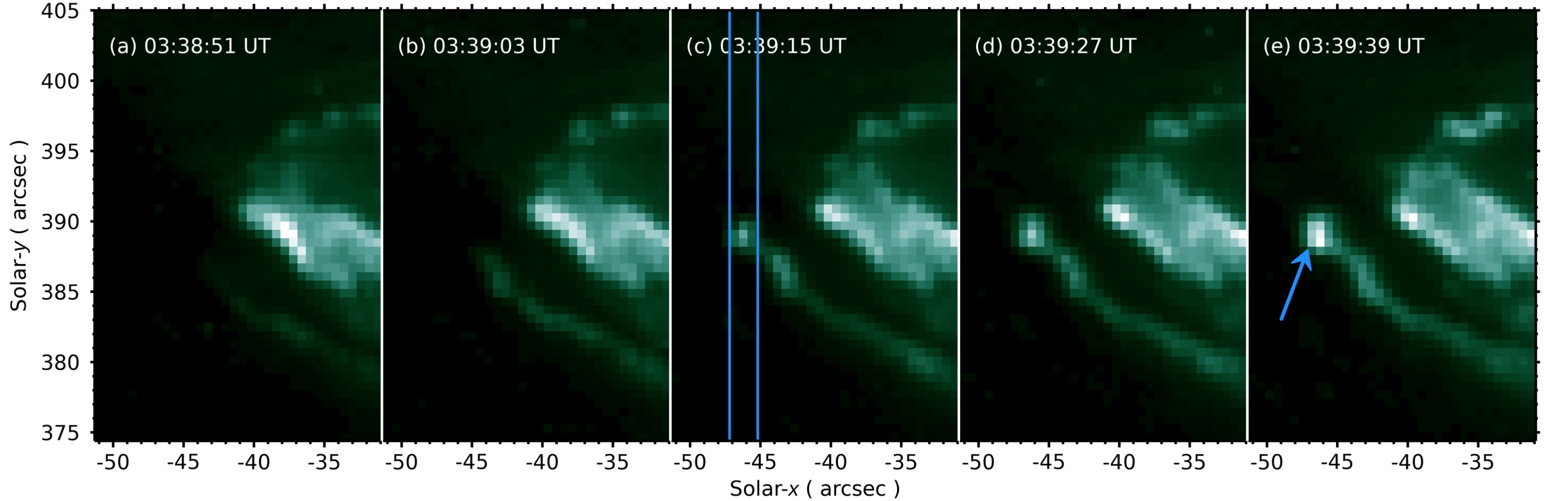}
    \caption{Five consecutive AIA 94~\AA\ images showing the development of the flare ribbon. A linear intensity scaling is used. Vertical lines on Panel (c) show the location of the EIS slit at the time the EIS spectrum was obtained. A blue arrow highlights the time of peak brightness of the flare kernel. }
    \label{fig.closeup}
\end{figure*}

Images from AIA are available for the flare, and Figure~\ref{fig.closeup} shows the rapid evolution of a section of the flare ribbon during a 1-min period as seen in the 94~\AA\ channel. This channel is the only one of the seven EUV channels that does not show saturation and so it best shows the rapid evolution of the flare ribbon. Normally in flare conditions this channel is dominated by the \ion{Fe}{xviii} 93.93\,\AA\ ($T_\mathrm{iz}=7$\,MK) line, but the EIS spectrum shows that lines formed at 4\,MK and higher are weak (see below). In this case the 94\,\AA\ channel emission mostly comes from \ion{Fe}{x} and \ion{Fe}{xiv} \citep{2012A&A...546A..97D}. 
Figure~\ref{fig.closeup} shows the rapid brightening of one kernel that reaches its peak brightness in Panel (e).  It is this feature that is observed by EIS and Panel (c) shows the location of the EIS slit at the time of this exposure.

\begin{figure*}
    \centering
    \includegraphics[width=\textwidth]{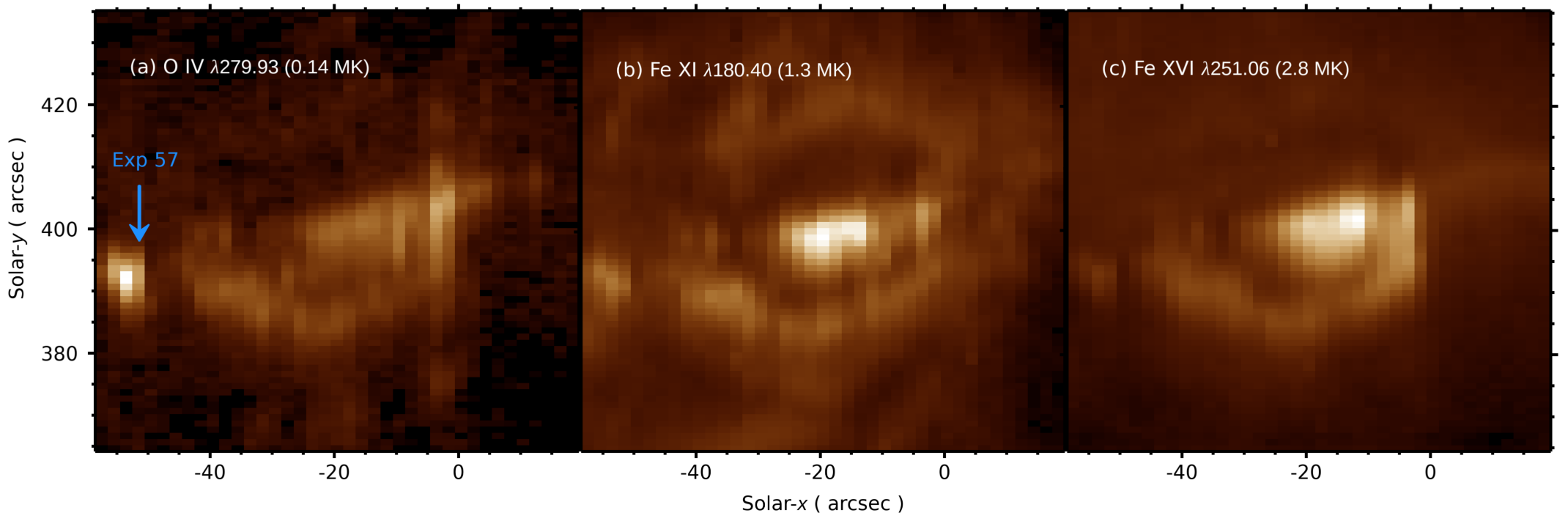}
    \caption{EIS raster images of AR 11429 in three emission lines. The temperature of formation of each ion is indicated, and the image column corresponding to exposure 57 is highlighted on Panel (a). A square-root intensity scaling is applied to each image.}
    \label{fig.eis-im}
\end{figure*}

EIS obtained  a single raster of the flaring region beginning at 03:09~UT and ending at 03:41~UT. Sixty exposures of 30~s were obtained with the 2\as\ slit and a step size of 2\as\ was used. The raster area was 120\as\ $\times$ 160\as. Figure~\ref{fig.eis-im} shows EIS raster images formed in three emission lines for a sub-region corresponding to exposures 22 to 60 (EIS rasters from right to left). The compact brightening at the left of Panel (a) corresponds to the flare kernel in Figure~\ref{fig.closeup}(e). To determine the spatial location of the EIS slit, synthetic raster images were generated from the AIA 94~\AA\ images through the procedure described in \citet{young_2023_10371938}. These images were compared with the image from the EIS \ion{Fe}{xiv} 274.20\,\AA\ line since \ion{Fe}{xiv} is one of the two major contributors to the AIA channel.\footnote{The other is \ion{Fe}{x} and the EIS \ion{Fe}{x}  184.54\,\AA\ image showed very similar morphology to the \ion{Fe}{xiv} image.} By adjusting the EIS slit in relation to the AIA frames it was possible to determine that the EIS slit was positioned as shown in Figure~\ref{fig.closeup}(c) at the time EIS exposure 57 was taken. This exposure began at 03:39:07~UT and completed at 03:39:37~UT. It thus captured the initial brightening of the kernel. The kernel continued to brighten after this exposure, when the EIS slit moved to the left by 2\as. Although the slit was not directly overlying the kernel for this exposure, the lower spatial resolution of EIS meant that the kernel still gave a strong signal, hence it is this exposure that is brightest in the EIS images (Figure~\ref{fig.eis-im}).

Figure~\ref{fig.eis-im} demonstrates that the flare kernel is observed at temperatures between 0.1 and 3~MK, although it is significantly brighter in \ion{O}{iv} in relation to the region at ($-20,-400$) compared to the hotter lines. The kernel is very weak or not visible in lines formed at 4~MK or higher, as found by considering \ion{Fe}{xvii} \lam204.67, \ion{Ca}{xv} \lam200.98 and \ion{Ca}{xvi} \lam208.59.

Exposure 57 was chosen rather than the brighter exposure 58 because the latter exhibited lines that are asymmetric, with an extended long-wavelength wing making spectral modeling more difficult. A spectrum averaged across the brightening in exposure 57 was obtained by first calibrating the level-0 EIS FITS file using the IDL routine \textsf{eis\_prep} with the standard options described in \citet{young_2022_7255646} plus the additional option \textsf{/nocr}. This prevented the routine from applying the cosmic ray removal procedure as it was found that the intense, compact nature of the  brightening led to spurious cosmic ray detections.  A pixel mask was then defined to select seven $y$-pixels in exposure 57 centered on the brightest pixel. The IDL routine \textsf{eis\_mask\_spectrum} was then run to create  1D spectra for the kernel for the SW and LW channels based on this mask.  Intensities quoted in this article are derived by summing the intensities of the seven individual pixels rather than averaging, as it is assumed the flare kernel is not spatially resolved by EIS. This is confirmed in Section~\ref{sect.diag} where the plasma emitting volume is estimated. There are two radiometric calibration options for EIS and the \href{https://solarb.mssl.ucl.ac.uk/JSPWiki/Wiki.jsp?page=EISAnalysisGuide}{EIS data analysis guide} does not recommend one over the other. One is due to \citet{2013A&A...555A..47D} and the other due to \citet{2014ApJS..213...11W}, which are referred to as GDZ13 and WUL14 in the remainder of this article. There are significant differences between the two that impact the \ion{O}{iv} diagnostics. The WUL14 calibration is preferred here and further details are given in Appendix~\ref{app.calib} and the following text.

The \textsf{/shift} option was used in the call to \textsf{eis\_mask\_spectrum}, which applies corrections due to the tilt of the EIS slit and the thermally-induced orbit variation of the spectrum on the detector. This leads to wavelengths that should be accurate to around 5~\kms\ \citep{2010SoPh..266..209K} for lines in the SW channel. There is an additional time-varying wavelength variation between the SW and LW channels \citep{2012ApJ...744...14Y} that was corrected by forcing \ion{Fe}{viii} \lam185.21 and \ion{Si}{vii} \lam275.37 to have the same Doppler shift, as recommended by  \citet{2012ApJ...744...14Y}.  Doppler velocities for the \ion{O}{iv} lines are given in Table~\ref{tbl.ints}, where they can be compared with other lines formed between 0.2 and 0.6~MK that are measured in the spectrum.

\citet{2013ApJ...767...55D} used the same EIS dataset to investigate chromospheric evaporation during the flare, concentrating on the bright emission around ($-20,-400$). They did note, however, that the feature at the left side of the raster  was brightest in transition region ions.

\section{Emission line measurements}\label{sect.meas}

Emission line intensities for the seven groups of \ion{O}{iv} lines listed in Table~\ref{tbl.trans} are derived through fitting Gaussian functions. Due to self-blending within the groups certain constraints are applied to the fits. For example, forcing the \ion{O}{iv} lines to have the same width and/or fixing the wavelength separations of the lines using the wavelengths from CHIANTI. The Gaussian fit parameters centroid, $\lambda_\mathrm{meas}$, full-width at half-maximum, $w$, and integrated intensity, $I$, are given in Table~\ref{tbl.ints}. The reference wavelengths, $\lambda_\mathrm{ref}$, are used to derive line-of-sight (LOS) Doppler velocities, $v_\mathrm{LOS}$. The uncertainties combine fitting uncertainties with radiometric calibration uncertainties (Appendix~\ref{app.calib}). The two EIS wavelength channels are discussed separately below.

In addition to the \ion{O}{iv} lines, a number of other emission lines are fit and presented in Table~\ref{tbl.ints}. These are either discussed in the following text or the lines are strong transition region lines for which the Doppler velocities can be compared with \ion{O}{iv}. 
It can be seen that the transition region line velocities range from $+20$ to $+45$\,\kms, with a trend of increasing redshift with ionization level. The coronal lines have smaller velocities, however.

\subsection{Lines in the LW channel}\label{sect.lw}

\begin{deluxetable}{ccclcc}[t]
\tablecaption{Gaussian fit parameters and identifications for selected emission lines.\label{tbl.ints}}
\tablehead{
    \colhead{$\lambda_\mathrm{meas}$} &
    \colhead{$w$} &
    \colhead{$I$} &
    \colhead{Ion} &
    \colhead{$\lambda_\mathrm{ref}$\tablenotemark{a}} &
    \colhead{$v_\mathrm{LOS}$} \\
    \colhead{(\AA)} &
    \colhead{(m\AA)} &
    \colhead{(\ecss)} &
    &
    \colhead{(\AA)} &
    \colhead{(\kms)} 
}
\tablecolumns{6}
\startdata
\sidehead{Transition region lines}
 184.143                  &     89 & $   14987 \pm  210$ &        \ion{O}{vi} &                   184.117 &                      41.7 \\
 185.240                  &     86 & $   61278 \pm  329$ &     \ion{Fe}{viii} &                   185.213 &                      44.2 \\
 192.927                  &     92 & $   19908 \pm  126$ &         \ion{O}{v} &                   192.904 &                      35.3 \\
 195.882                  &     78 & $     896 \pm   70$ &        \ion{O}{iv} &                   195.860 &                      33.7 \\
 207.251                  &     65 & $    1526 \pm  126$ &        \ion{O}{iv} &                   207.239 &                      18.1 \\
 248.480\tablenotemark{b} &    102 & $   10990 \pm  238$ &         \ion{O}{v} &                   248.460 &                      24.0 \\
 249.154                  &    119 & $   18319 \pm  294$ &       \ion{Si}{vi} &                   249.124 &                      35.7 \\
 250.157                  &     97 & $    1729 \pm   91$ &     \ion{Al}{viii} &                   250.134 &                      27.3 \\
 268.046                  &    102 & $     357 \pm   35$ &        \ion{O}{iv} &                   268.024 &                      24.2 \\
 269.019                  &    105 & $   15813 \pm  798$ &       \ion{Mg}{vi} &                   268.991 &                      31.7 \\
 272.150                  &    106 & $    7483 \pm  385$ &        \ion{O}{iv} &                   272.127 &                      25.6 \\
 275.409                  &    118 & $   20811 \pm 1050$ &      \ion{Si}{vii} &  275.368\tablenotemark{c} &     44.2\tablenotemark{d} \\
 276.178                  &    103 & $    5761 \pm  301$ &      \ion{Mg}{vii} &  276.142\tablenotemark{c} &                      39.1 \\
 276.613                  &     97 & $   24171 \pm 1218$ &        \ion{Mg}{v} &                   276.579 &                      36.3 \\
 279.656                  &     89 & $    4977 \pm  322$ &        \ion{O}{iv} &                   279.631 &                      27.1 \\
 279.959                  &     96 & $   10941 \pm  700$ &        \ion{O}{iv} &                   279.933 &                      28.2 \\
 280.769                  &    106 & $   24766 \pm  210$ &      \ion{Mg}{vii} &  280.729\tablenotemark{c} &                      43.0 \\
\noalign{\smallskip}
\sidehead{Coronal lines}
 251.956                  &    124 & $   15190 \pm  224$ &     \ion{Fe}{xiii} &                   251.952 &                       5.1 \\
 261.067                  &    119 & $    7294 \pm   98$ &        \ion{Si}{x} &                   261.056 &                      12.6 \\
 272.009                  &    138 & $    8694 \pm  462$ &        \ion{Si}{x} &                   271.992 &                      18.3 \\
\enddata
\tablenotetext{a}{From CHIANTI unless otherwise indicated.}
\tablenotetext{b}{Includes contribution of 147\,\ecss\ from \ion{Al}{viii} \lam284.45 (Section~\ref{sect.comp}).}
\tablenotetext{c}{\citet{2023ApJ...958...40Y}.}
\tablenotetext{d}{Velocity forced to be the same as \ion{Fe}{viii} \lam185.21 (Section~\ref{sect.obs}).}
\end{deluxetable}

\begin{figure*}[t]
    \centering
    \includegraphics[width=\textwidth]{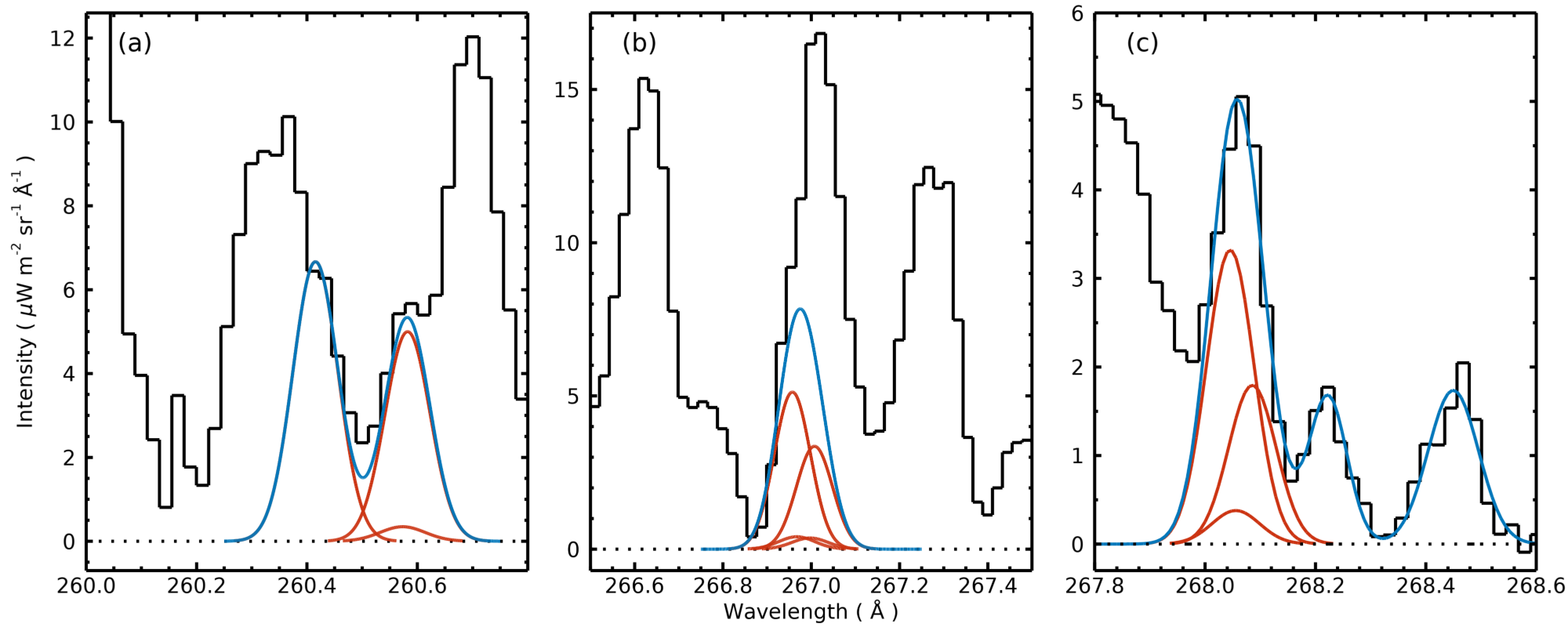}
    \caption{Portions of the flare spectrum (black) showing the \ion{O}{iv} lines near 260.5~\AA\ (a), 267.0~\AA\ (b), and 268.0~\AA\ (c), corresponding to the TA6a,b and TA7 transitions, respectively. Blue  curves for Panels (a) and (b) show the estimated total line profiles and red curves show individual \ion{O}{iv} components. The blue curve on Panel (c) shows the total multi-Gaussian fit function, and the red curves show the individual \ion{O}{iv} components of the TA7 line group.}
    \label{fig.260}
\end{figure*}

The two lines from TA6a at 260.39~\AA\ and 260.56~\AA\ are weak and partly blended with other species. 
In addition, the presence of multiple lines close together through this part of the spectrum makes estimating the spectrum background level difficult. Rather than attempt the fitting of multiple Gaussian functions, the line intensities are estimated by simply over-plotting the Gaussian fit function of the stronger 279.93~\AA\ line on the spectra at the locations of the lines, as shown in Figure~\ref{fig.260}(a). The TA6a lines are assumed to have the same velocity as \lam279.93. The relative strengths of the TA3a lines are set to the ratios given in CHIANTI, and their absolute magnitude has been set to 1/16 of the 279.93~\AA\ line. The continuum level was estimated from a line-free region outside of the displayed wavelength region. It is clear that the two principal TA6a lines are a good match for two features in the spectra, although both are partly blended with nearby stronger lines. The feature at 260.3~\AA\ is unknown, but that at 260.7~\AA\ is \ion{Fe}{vii} 260.67~\AA\ \citep{2009ApJ...707..173Y}. Based on a by-eye comparison of estimated profiles and the observed spectrum, the \lam279.93/\lam260.39 ratio is estimated to be $16\pm 2$. This value is discussed further in Section~\ref{sect.diag}.

The TA6b lines at 266.93~\AA\ and 266.98~\AA\ (Table~\ref{tbl.trans}) are  blended with each other at the EIS resolution, and a line is clearly seen at this location (Figure~\ref{fig.260}(b)). However, it is too strong to be explained by the \ion{O}{iv} lines, as illustrated by the over-plotted blue profile on Figure~\ref{fig.260}(b). As above, the profiles are Gaussians with the same width as \lam279.93 but shifted to the expected locations of the TA6b lines using the CHIANTI wavelengths. Relative amplitudes between the TA6b lines are set by the CHIANTI theoretical ratios, and the absolute magnitude is set by using the CHIANTI \lam279.93/\lam266.93 ratio of 20.8 obtained at $\log\,(N_\mathrm{e}/\mathrm{cm}^{-3})=12.0$ and $\log\,(T/\mathrm{K})=5.10$ (Section~\ref{sect.diag}).
As the TA6b lines can not be reliably separated from the stronger feature, they are not used in the present analysis.

A line at 268~\AA\ corresponds to a TA7 multiplet that is also blended at the EIS resolution. Measurement of this feature is complicated by a broad spectral feature to the short-wavelength side, and a weak line on the long-wavelength side. A multi-Gaussian fit was performed, and the result is shown in Figure~\ref{fig.260}(c). The three \ion{O}{iv} lines (shown in red) were fit by forcing the two weaker lines to have fixed wavelength separations relative to the stronger line. Widths were forced to be equal, and the amplitudes were fixed to the ratios predicted by CHIANTI. The two weaker, unknown lines at longer wavelengths were free to vary and the background was forced to be flat.

\begin{figure*}
    \centering
    \includegraphics[width=\textwidth]{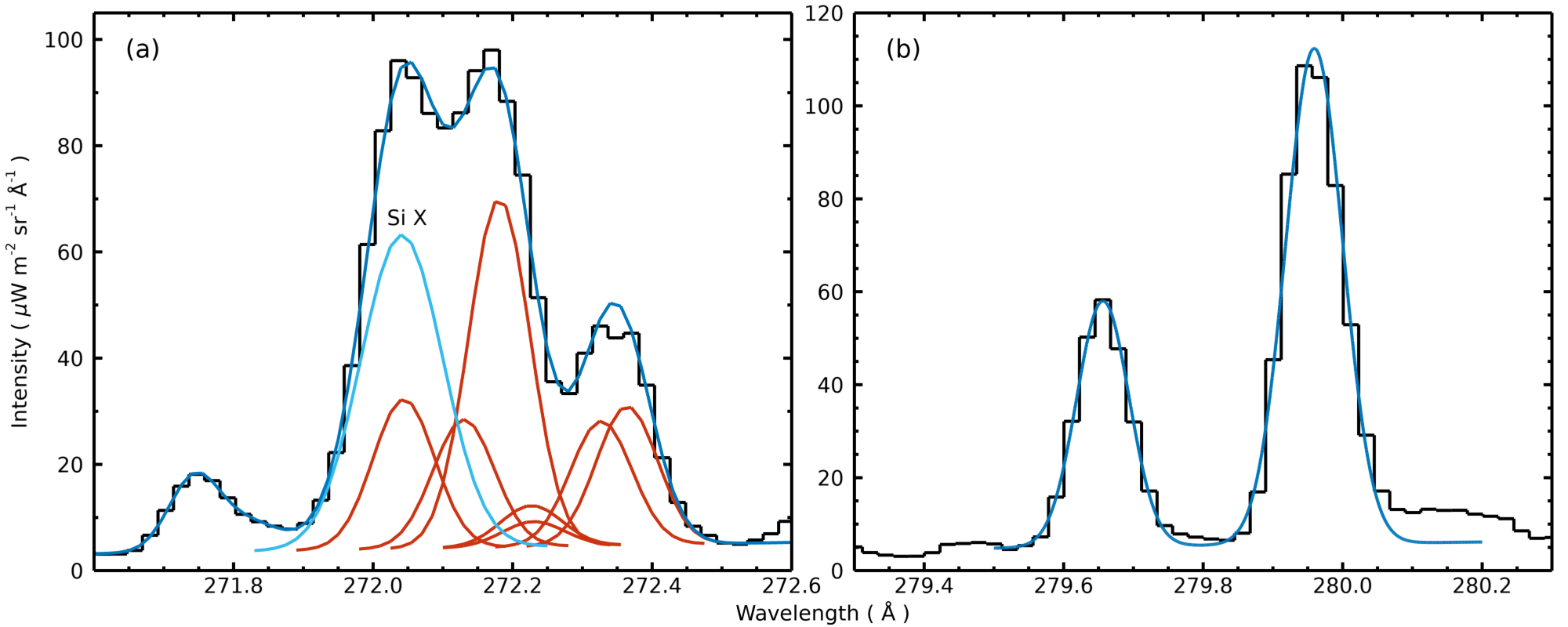}
    \caption{(a) The spectral region containing the \ion{O}{iv} multiplet around 272~\AA. The multi-Gaussian fit to the entire spectral feature is plotted in blue, the seven individual \ion{O}{iv} components are shown in red, and \ion{Si}{x} \lam271.99 is shown in cyan. (b) The spectral region containing \ion{O}{iv} \lam\lam279.63,279.93, with the two-Gaussian fit to the lines over-plotted in blue.}
    \label{fig.272-279}
\end{figure*}

The seven \ion{O}{iv} lines around 272~\AA\ are part of a complex spectral feature (Figure~\ref{fig.272-279}(a)) that includes \ion{Si}{x} \lam271.99, which is dominant in most solar conditions. The intensity of the complete feature is around 4100\,\ecss. There is an \ion{Fe}{xiii} line expected at 272.19~\AA, but this can be neglected: the \ion{Fe}{xiii} \lam272.19/\lam251.96 ratio is predicted from CHIANTI to have a fixed ratio of 0.019, and the measured  \lam251.96 intensity (Table~\ref{tbl.ints}) implies a \lam272.19 intensity of only  41~\ecss, so only 1\,\%\ of the total feature.
A highly-constrained multi-Gaussian fit was performed to estimate the \ion{O}{iv} intensities.
Since the seven \ion{O}{iv} lines show little temperature or density sensitivity relative to each other, then the parameters of six of the lines were tied to those of the strongest line at 272.13~\AA. CHIANTI yields the predicted ratios of all the lines relative to \lam272.13, and their wavelength offsets. Thus the peaks of the six lines were forced to be fixed ratios relative to \lam272.13, the centroids were forced to have fixed offsets relative to \lam272.13 and the widths were forced to be the same as \lam272.13. Three additional Gaussians were added to the fit function, representing \ion{Si}{x} \lam271.99, \ion{Fe}{vii} \lam271.69 and the unknown line at about 271.85~\AA. The spectrum background was fit with a linear function.
The \textsf{auto\_fit} and \textsf{mask\_spectrum} suites of routines \citep{young_2022_6339584} were used to prepare the data, set up the template and perform the fit. The results are shown in Figure~\ref{fig.272-279}(a) where the seven \ion{O}{iv} lines are displayed as red curves, and the line fit parameters of \lam272.13 line  are given in Table~\ref{tbl.ints}. 
The combined intensity for all of the \ion{O}{iv} lines is $2842\pm 216$~\ecss, although we only consider the \lam272.13 intensity for the remainder of this article.
The width and Doppler velocity for \lam272.13 are in good agreement with the other LW \ion{O}{iv} lines in Table~\ref{tbl.ints}, giving confidence in the fit.

\ion{Si}{x} \lam271.99 is shown as a cyan curve in Figure~\ref{fig.272-279}(a), and the fit parameters are given in Table~\ref{tbl.ints}. Also given in this table are the fit parameters for  \ion{Si}{x} \lam261.06, which is isolated in the spectrum and fit with a single Gaussian. The \lam261.06/\lam271.99 ratio is insensitive to density with an empirical value from off-limb quiet Sun spectra of 0.97 using the WUL14 calibration. The ratio from the flare spectrum is 0.84, lower than expected, and the \lam271.99 width is larger and the redshift larger. This suggests the multi-Gaussian fit for the complex feature at 272~\AA\ is over-estimating the \ion{Si}{x} line. This could be due to the \ion{O}{iv} intensities being under-estimated by around 5\%, but also could be due to an unknown blend close to the \ion{Si}{x} line. As a 5\,\%\ uncertainty was added to the \ion{O}{iv} intensity measurements (Appendix~\ref{app.calib}), the multi-Gaussian fit to the \ion{O}{iv} lines is retained.

The two lines at 279~\AA\ are isolated in the spectrum, however the background level in the spectrum is difficult to estimate due to weak features neighboring the lines. Fits were performed for two extreme choices of the spectrum background, and for a compromise estimate of the background. The latter fit is shown in Figure~\ref{fig.272-279}(b), and the fit parameters are given in Table~\ref{tbl.ints}. The uncertainties on the integrated intensities have been increased by 4\%\ of the measured intensities (added in quadrature), based on comparisons with the two extreme fits.
The \lam279.63/\lam279.93 intensity ratio is $0.455\pm 0.026$ which is slightly lower than the expected ratio of 0.499 from CHIANTI.

\subsection{Lines in the SW channel}\label{sect.sw}

The large uncertainties for the SW--LW cross-calibration as inferred from the GDZ13 and WUL14 calibrations (Appendix~\ref{app.calib}) mean that the SW \ion{O}{iv} lines can not be reliably used in diagnostics with the LW lines. However, in this section the SW lines are measured and in Section~\ref{sect.diag} they are assessed in terms of the plasma parameters inferred from the LW channel lines.

\begin{figure*}[t]
    \centering
    \includegraphics[width=\textwidth]{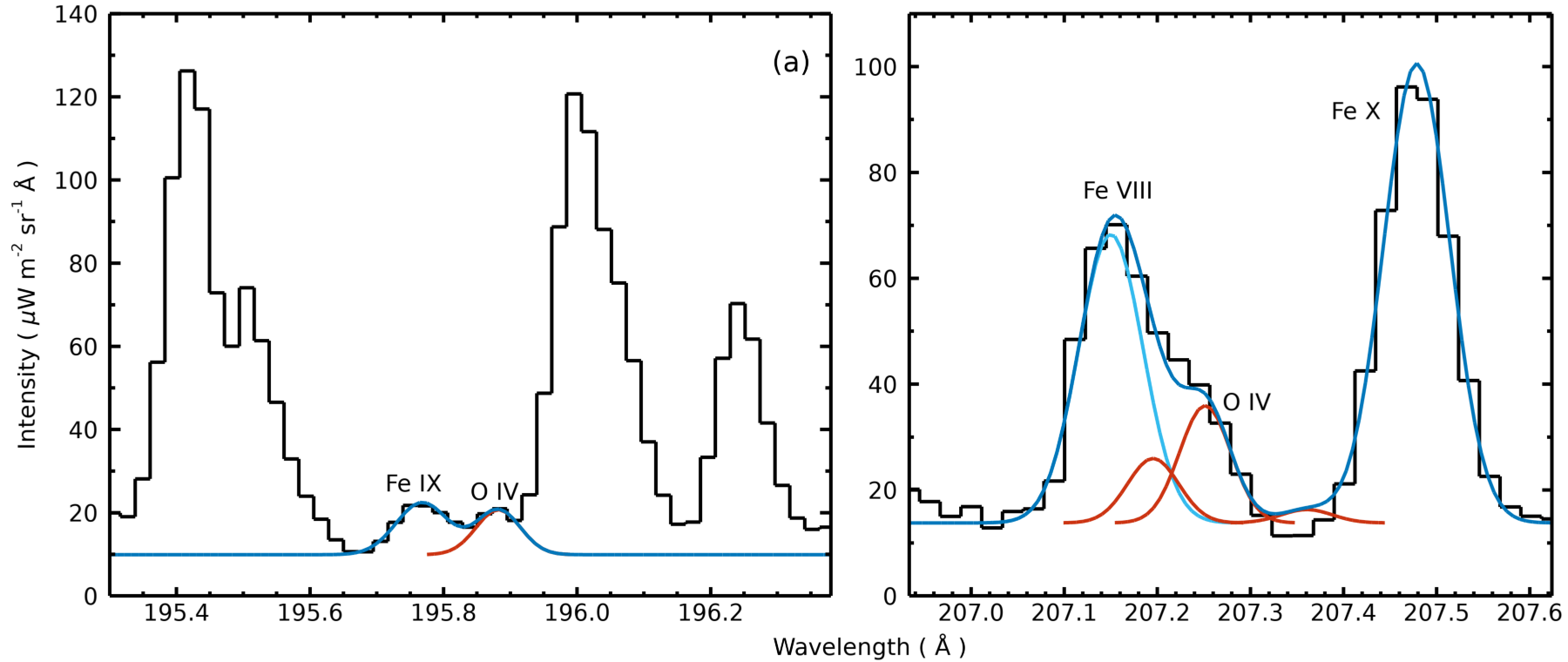}
    \caption{(a) The spectral region containing the weak \ion{O}{iv} 195.86~\AA\ line. A two-Gaussian fit is shown in blue with the \ion{O}{iv} line represented by the red curve. (b) The spectral region containing the \ion{O}{iv} TA4 lines, with a five-Gaussian fit shown in blue. The three \ion{O}{iv} components are shown in red.}
    \label{fig.207}
\end{figure*}

Two lines from TA3 are expected at 195.86~\AA\ and 196.01~\AA, with an expected ratio of 1:1.8. Figure~\ref{fig.207}(a) shows this region of the spectrum, which is dominated by lines of \ion{Fe}{vii} and \ion{Fe}{viii} \citep{2021ApJ...908..104Y}. 
The stronger of the two \ion{O}{iv} lines is blended with \ion{Fe}{viii} 195.97~\AA, which is stronger in all solar conditions including the flare kernel considered here. A weak feature is seen at the expected location of the 195.86~\AA\ line, and a two-Gaussian fit was performed to fit this feature and an \ion{Fe}{ix} line of comparable intensity at shorter wavelengths. The fit is shown in Figure~\ref{fig.207}(a)  and the \ion{O}{iv} fit parameters are given in Table~\ref{tbl.ints}. The Doppler velocity is in reasonable agreement with the LW channel lines. The main uncertainty (not contained in the formal fitting uncertainty in Table~\ref{tbl.ints}) is the spectrum background level, which is not well constrained in this region.

TA4 has two lines at 207.18 and 207.24~\AA\ that have a predicted ratio of 1:1.8. \citet{2010ApJ...711...75L} gave an intensity measurement for \lam207.24
for an off-limb coronal mass ejection observation with EIS, and used the \lam207.24/\lam279.93 ratio to obtain lower limits on the density at two times during the event. 
In the present spectrum, and also the loop footpoint spectra of \citet{2009ApJ...706....1L} and \citet{2009A&A...508..501D} there is a line at 207.15~\AA\ due to \ion{Fe}{viii}, with \ion{O}{iv} responsible for the weaker emission on the long-wavelength side of this line (Figure~\ref{fig.207}(b)). 
A multi-Gaussian fit was performed to lines in this region, and is shown as the blue curve on Figure~\ref{fig.207}(b).
The three \ion{O}{iv} lines are shown in red, and each of the three Gaussian parameters of \lam207.18 and \lam207.35 were tied to those of \lam207.24. In particular, the peaks were set to factors 0.55 and 0.114 of the peak of \lam207.24 based on the CHIANTI emissivity model, and the centroid offsets were fixed based on the CHIANTI wavelengths. The fit parameters for \lam207.24 are given in Table~\ref{tbl.ints}, and the Doppler velocity is in reasonable agreement with the other \ion{O}{iv} lines, but the width is much narrower. The reduced $\chi^2$ value for the fit is rather high at 6.2, and there is no evidence for the weak \ion{O}{iv} line at 207.35~\AA\ in the spectrum (Figure~\ref{fig.207}(b)).

\section{Diagnostics}\label{sect.diag}

Figure~\ref{fig.dens} shows the temperature and density derived from the \lam268.02/\lam279.93 and \lam272.13/\lam279.93 ratios. The temperature is $\log\,(T/\mathrm{K})=5.10\pm 0.03$, and the density is $\log\,(N_\mathrm{e}/\mathrm{cm}^{-3})= 12.55$ with a lower limit of 11.91. The upper limit of the measured ratio is seen to be above the maximum value of theoretical ratio. However, the theoretical ratio reaches a maximum at $\log\,(N_\mathrm{e}/\mathrm{cm}^{-3})= 13$ and then decreases, thus the lower limit of the measured ratio also yields an upper limit to the density of $\log\,(N_\mathrm{e}/\mathrm{cm}^{-3})\le 14.40$.

The derived temperature is surprisingly low. The contribution function of \lam279.93  peaks at $\log\,(T/\mathrm{K})=5.26$. When convolved with the flare differential emission measure (DEM) curve distributed with CHIANTI, the function peaks at $\log\,(T/\mathrm{K})=5.20$ as the DEM increases with decreasing temperature. CHIANTI has the option to compute a revised ionization equilibrium with  density-dependent suppression factors for dielectronic recombination \citep{2021ApJ...909...38D}. This has been demonstrated to shift the formation temperatures of oxygen ions to lower temperatures \citep{2018ApJ...857....5Y}. Computing the \ion{O}{iv} fractions using DR suppression at $\log\,(N_\mathrm{e}/\mathrm{cm}^{-3})=12.0$ gives a \lam279.93 contribution function that peaks at $\log\,(T/\mathrm{K})=5.18$, and convolving with the CHIANTI flare DEM gives a peak at $\log\,(T/\mathrm{K})=5.13$. \citet{2020MNRAS.497.1443D} found that including level-resolved ionization and recombination in the ion balance calculations serves to further lower the formation temperature of \ion{O}{iv} by 0.05~dex. The results from the EIS temperature diagnostic discussed in this article therefore gives support to the addition of density-dependent factors to the standard CHIANTI ionization balance calculation.

From the density and temperature an estimate of the size of the flare kernel at \ion{O}{iv} temperatures can be made. Assuming an isothermal plasma, the  column depth, $h$, is given by
\begin{equation}
    h= {4\pi I \over 0.83 \epsilon(\mathrm{O)}G(T,N_\mathrm{e}) N^2_\mathrm{e} }
\end{equation}
where $\epsilon(\mathrm{O})$ is the abundance of oxygen relative to hydrogen, and $G$ is the contribution function. The contribution function is calculated from the CHIANTI ionization balance calculated at an electron pressure of $10^{17.1}$\,K\,cm$^{-3}$, and a temperature of $\log\,(T/\mathrm{K})=5.10$. The density is set to $\log\,(N_\mathrm{e}/\mathrm{cm}^{-3})= 12$ to be consistent with the assumed pressure. The photospheric abundance file distributed with CHIANTI then gives $h=0.036$\as. The EIS pixel size is 2\as\ $\times$ 1\as, so if a cubic volume is assumed then the side of this cube is 0.4\as\ (300\,km). This compares with directly measured flare ribbon sizes of 0.6\as\ to 2.0\as\ obtained from the Goode Solar Telescope at chromospheric temperatures \citep{2012ApJ...750L...7X}. Due to the uncertainties in the density and the element abundance, the EIS size estimate is perhaps accurate to no better than a factor of two. 

Section~\ref{sect.lw} discussed the \ion{O}{iv} lines near 260~\AA\ and an estimate for the \lam279.93/\lam206.39 ratio was given as $16\pm 2$. The ratio is strongly temperature sensitive with little density sensitivity, and the observed ratio corresponds to $\log\,(T/\mathrm{K})=5.07\pm 0.03$, consistent with the value from \lam268.02/\lam279.93. 

The EIS SW lines at 195.86~\AA\ and 207.24~\AA\ form a ratio that is sensitive to both temperature and density, although much less so than the \lam268.02/\lam279.93 and \lam272.13/\lam279.93 ratios. For a temperature of $\log\,(T/\mathrm{K})=5.10$, the measured ratio of $1.70\pm 0.19$ (Table~\ref{tbl.ints}) implies a lower limit to the density of $\log\,(N_\mathrm{e}/\mathrm{cm}^{-3})\ge 11.85$, which is consistent with the LW diagnostics discussed above. This gives confidence in the accuracy of the intensity measurements of these lines, despite the issues discussed in Section~\ref{sect.sw}.

The \lam207.24/\lam279.93 ratio is strongly temperature sensitive and also has weak density sensitive. Assuming $\log\,(N_\mathrm{e}/\mathrm{cm}^{-3})= 12.55$, the observed ratio of $0.139\pm 0.015$ implies a temperature of $\log\,(T/\mathrm{K})=5.12\pm 0.04$. Decreasing the density to $\log\,(N_\mathrm{e}/\mathrm{cm}^{-3})= 11.91$ increases the temperature by 0.01~dex. The ratio is therefore consistent with that derived from the \lam268.02/\lam279.93 ratio, and gives confidence in the WUL14 calibration.

\section{Comparison with other density diagnostics}\label{sect.comp}

The closest ion to \ion{O}{iv} ($T_\mathrm{iz}=0.14$~MK) in temperature that also has a density diagnostic is \ion{O}{v} ($T_\mathrm{iz}=0.22$~MK). A group of lines between 192.7 and 193.0~\AA\ is density sensitive relative to \lam248.46 \citep{2007PASJ...59S.857Y}, with the former becoming relatively stronger at higher densities. The ratio is sensitive to densities up to $10^{13}$~cm$^{-3}$ and thus potentially provides confirmation of the high \ion{O}{iv} density.

There are two complicating factors for the \ion{O}{v} ratio: (i) the lines are found in the two different EIS wavelength channels; and (ii) the lines are affected by blending. Appendix~\ref{app.calib} shows that the GDZ13 calibration gives a ratio 36\,\%\ lower than for the WUL14 calibration, which is preferred in this article. The effect on the derived density is discussed below.

\begin{figure*}[t]
    \centering
    \includegraphics[width=0.9\textwidth]{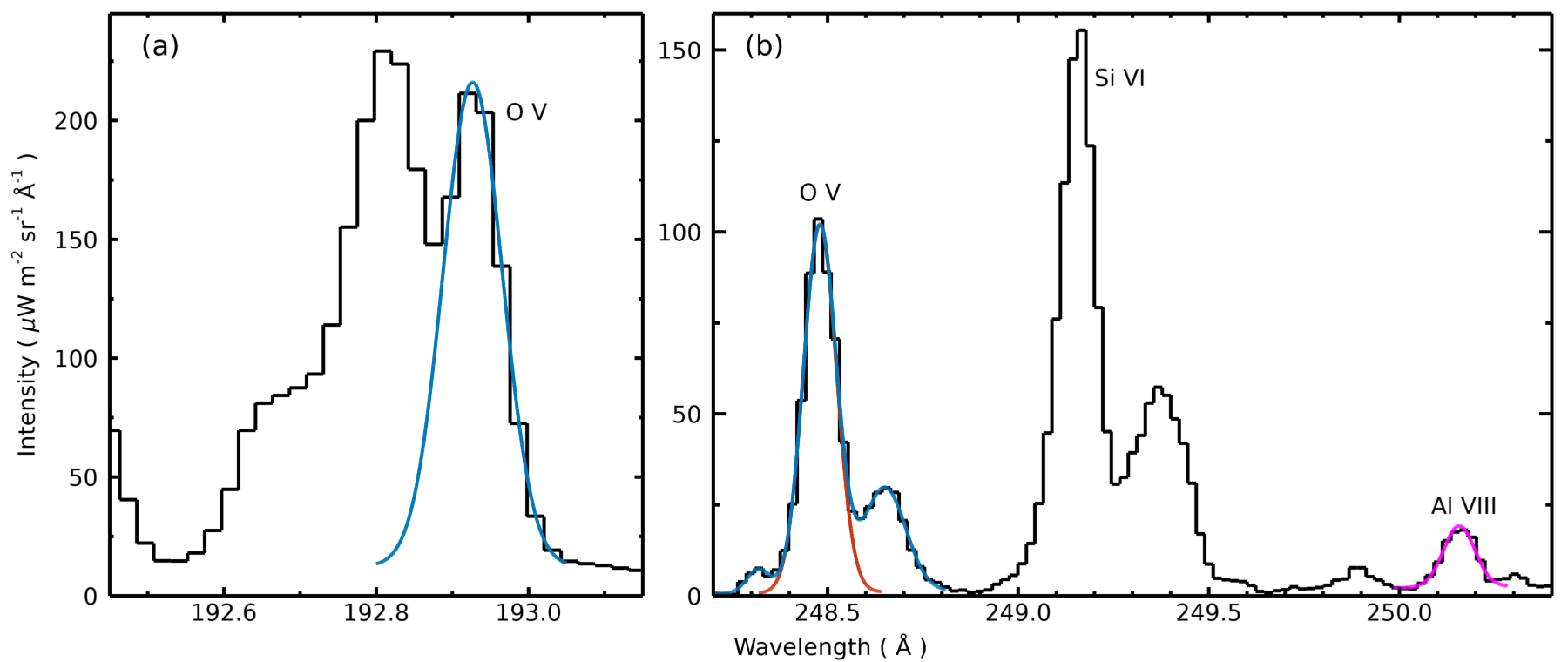}
    \caption{Portions of the EIS flare kernel spectra showing the \ion{O}{v} lines. (a) the fit to \ion{O}{v} \lam192.90 is indicated in blue. (b) a three-Gaussian fit (blue) was performed to the region containing \ion{O}{v} \lam248.46 (red), and \ion{Al}{viii} \lam250.13 is fit with a single Gaussian (magenta).]}
    \label{fig.o5}
\end{figure*}

Blending issues associated with the group of lines between 192.7 and 193.0~\AA\ have been discussed by \citet{2009ApJ...697.1956K}. In particular, \ion{Fe}{xi} \lam192.81 is usually dominant in quiet Sun and active region conditions, while \ion{Ca}{xvii} 192.85 usually dominates in flare conditions. Rather than attempting the complex multi-Gaussian fit performed by \citet{2009ApJ...697.1956K} the procedure here is simply to fit the rightmost feature in the spectrum (Figure~\ref{fig.o5}) and assume this corresponds to the \ion{O}{v} self-blend of \lam192.904 and \lam192.911 (the latter contributes less than 15\%\ to the observed feature). This is justified because of the strength of \ion{O}{v} in the flare kernel spectrum, enabling \ion{O}{v} to be clearly separated from the other lines around 192.8~\AA. The line parameters given in Table~\ref{tbl.ints} are identified only with the 192.904~\AA\ line, but for computing the density both components are included in the CHIANTI model.

The \ion{O}{v} line at 248.46~\AA\ is blended with the \ion{Al}{viii} 248.45~\AA\ line, but the latter can be estimated by measuring \ion{Al}{viii} 250.13~\AA. The branching ratio \lam248.45/\lam250.13 has a fixed value of 0.594 in all conditions using atomic data in CHIANTI. Fits to the lines are shown in Figure~\ref{fig.o5}(b), and the \lam250.13 intensity is given in Table~\ref{tbl.ints}. It implies a \ion{Al}{viii} \lam248.45 intensity of 147~\ecss, and the \ion{O}{v} \lam248.46 intensity is then $1423\pm 35$~\ecss. 

The measured \lam192.90/\lam248.46 ratio is 2.00 with the WUL14 calibration, which translates to a density of $\log\,(N_\mathrm{e}/\mathrm{cm}^{-3})=11.43$. With the GDZ13 calibration, the ratio is only 1.28, corresponding to $\log\,(N_\mathrm{e}/\mathrm{cm}^{-3})=10.80$. The WUL14 calibration was preferred for the \ion{O}{iv} analysis (Appendix~\ref{app.calib}), and if an uncertainty of 20\%\ is assumed for the SW-LW cross-calibration, then this implies a density of $\log\,(N_\mathrm{e}/\mathrm{cm}^{-3})=11.4\pm 0.3$. This density was derived assuming a temperature of $\log\,(T/\mathrm{K})=5.35$, but the result is insensitive to temperature within $\pm 0.2$~dex.

\ion{Mg}{vii} ($T_\mathrm{iz}=0.63$~MK) is significantly hotter than \ion{O}{iv} but has density diagnostics that have been widely used \citep{2009ApJ...694.1256T,2012ApJ...744...14Y,2016ApJ...830..101B}. The \lam280.73/\lam276.15 observed ratio (Table~\ref{tbl.ints}) gives a density close to the high density limit: $\log\,(N_\mathrm{e}/\mathrm{cm}^{-3})=11.04^{+0.37}_{-0.20}$, computed at $\log\,(T/\mathrm{K})=5.80$.

\begin{deluxetable}{cccc}[t]
\tablecaption{Electron pressures for three ions.\label{tbl.press}}
\tablehead{
    \colhead{Ion} &
    \colhead{$\log\,(T_\mathrm{iz}/\mathrm{K})$} &
    \colhead{$\log\,(N_\mathrm{e}/\mathrm{cm}^{-3})$}  &
    \colhead{$\log\,(P_\mathrm{e}/\mathrm{K\,cm}^{-3})$} 
}
\startdata
\ion{O}{iv} & 5.10  & $\ge 11.9$  & $\ge 17.0$ \\
\ion{O}{v}  & 5.30 & $11.4\pm 0.3$ & $16.7 \pm 0.3$ \\
\ion{Mg}{vii} & 5.75 & $11.0^{+0.4}_{-0.2}$ & $16.8^{+0.4}_{-0.2}$ \\
\enddata
\end{deluxetable}

Table~\ref{tbl.press} compares the density results from the three ions and gives the electron pressure ($P_\mathrm{e}=N_\mathrm{e}T$). The $T_\mathrm{iz}$ values have been calculated using the CHIANTI ionization balance computed at a pressure of $\log\,(P_\mathrm{e}/\mathrm{K\,cm}^{-3})=17.1$ although the values do not change for pressures up to two orders of magnitude lower than this. The results are therefore consistent, within the uncertainties, with a pressure of $10^{17}$\,K\,cm$^{-3}$. For comparison, \citet{1987ApJ...320..913W} found that a number of density diagnostics, including \ion{O}{iv} and \ion{O}{v}, observed with the Skylab S082B spectrometer were consistent with a pressure of $3.9\times 10^{16}$\,K\,cm$^{-3}$ for a flare kernel observed at the solar limb.

\section{Comparison With Model Results}\label{sect.fchroma}

The Flare Chromospheres: Observations, Models and Archives (F-CHROMA) collaboration provided a number of flare model runs produced with the RADYN code.\footnote{\url{https://star.pst.qub.ac.uk/wiki/public/solarmodels/start.html}.} The models can be used to investigate how the density and temperature derived from the \ion{O}{iv} diagnostics can help constrain the flare heating process. The models give the electron density and temperature as a function of time, $t$, and position, $z$, along the flare loop. Three parameters are varied amongst the models: the spectral index, $\delta$, the low-energy cutoff, $E_\mathrm{c}$, and the total energy input, $E_\mathrm{tot}$. Full details are given in \citet{2023A&A...673A.150C}.

\begin{figure}[t]
    \centering
    \includegraphics[width=0.6\textwidth]{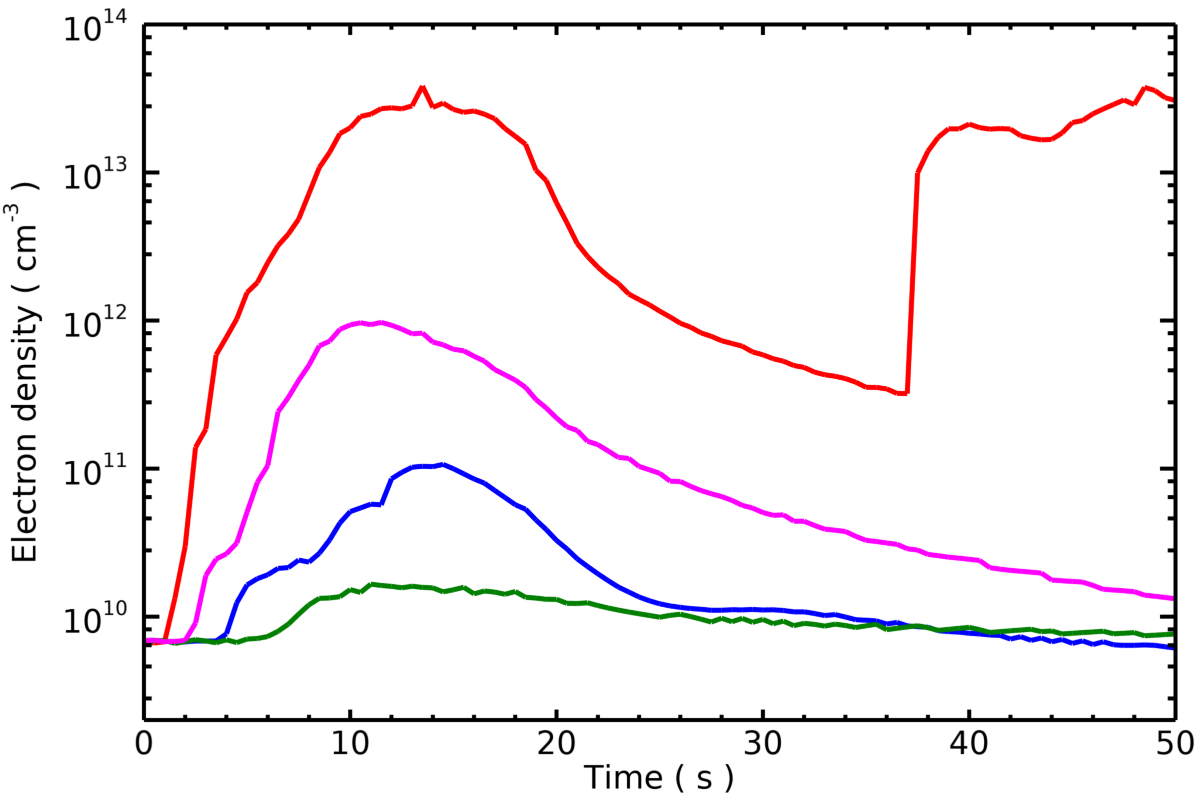}
    \caption{Plots showing the variation of electron density as a function of time for plasma at $\log\,(T/\mathrm{K})=5.1$ obtained from the F-CHROMA flare models. The colors green, blue, magenta and red correspond to $E_\mathrm{tot}=3\times 10^{10}$, $10^{11}$, $3\times 10^{11}$, and $10^{12}$\,erg\,cm$^{-2}$. For each model $\delta=4$ and $E_\mathrm{c}=20$\,keV.}
    \label{fig.model}
\end{figure}

The values $\delta=4$ and $E_\mathrm{c}=20$\,keV were chosen to represent a relatively strong flare, and there are four models with $E_\mathrm{tot}=3\times 10^{10}$, $10^{11}$, $3\times 10^{11}$, and $10^{12}$\,erg\,cm$^{-2}$. The data were degraded to 0.5\,s time resolution and for each time step the spatial pixels corresponding to the temperature region $\log\,T=5.0$ to 5.2 (spanning the temperature found from the \ion{O}{iv} diagnostic) were identified. An average density for each time step is computed as $\sum_i N_{\mathrm{e},i} z_i/\sum_i z_i$, where $z_i$ is the size of spatial bin $i$. Figure~\ref{fig.model} shows the variation of the average density as a function of time for each model. 
As can be seen, the maximum density reached at times $t=10$--15\,s increases strongly as $E_\mathrm{tot}$ increases. The rise in density at 37\,s for the $E_\mathrm{tot}=10^{12}$\,erg\,cm$^{-2}$ model is due to downward-propagating wave that has reflected from the loop top.

These simulations are not directly tuned to the present observations, and so a direct comparison between the results and the \ion{O}{iv} measurements can not be made. However, it demonstrates that the \ion{O}{iv} plasma measurements may be a valuable tool in constraining flare loop models. In particular, high densities of $\log\,(N_\mathrm{e}/\mathrm{cm}^{-3})\ge 12$, suggest a high energy flux into the flare loop.

\section{Summary}\label{sect.summary}

New \ion{O}{iv} diagnostics for temperature and density have been presented that utilize emission lines observed by Hinode/EIS. They have been applied to the spectrum of a solar flare kernel observed on 2012 March 9. The temperature is strongly constrained to $\log\,(T/\mathrm{K})=5.10\pm 0.03$ by the \lam268.02/\lam279.93 ratio. This temperature is significantly below the temperature given by the default CHIANTI ionization equilibrium calculation. However, recent work on density dependent ionization balance calculations \citep{2020MNRAS.497.1443D} places \ion{O}{iv} at lower temperatures in high density conditions, giving consistency with the flare temperature found here.

The measured \lam272.13/\lam279.93 ratio is close to the high density limit, and a lower limit of $\log\,(N_\mathrm{e}/\mathrm{cm}^{-3})=11.9$ is found here. Comparisons with \ion{O}{v} and \ion{Mg}{vii} density diagnostics suggest a constant pressure of $10^{17}$\,K\,cm$^{-3}$ could apply through the transition region if the \ion{O}{iv} density is near the lower density limit.
Assuming a density of $\log\,(N_\mathrm{e}/\mathrm{cm}^{-3})=12.0$ and a cubic emitting volume, the width of the volume is estimated at 0.4\as\ (300\,km).

Inspection of F-CHROMA flare loop models (Section~\ref{sect.fchroma}) shows that the highest densities reached during the flare heating process are strongly dependent on the total energy input to the loop. Although the flare models are not directly applicable to the present flare, they suggest that the high \ion{O}{iv} density found here is consistent with a high energy input to the flare loop footpoint. 

The  upcoming NASA Multi-slit Solar Explorer \citep[MUSE:][]{2020ApJ...888....3D} mission will perform spectroscopy in three EUV channels, one of which  is centered at 284~\AA. This channel will have sensitivity to the \ion{O}{iv} lines between 268 and 280~\AA---see Figure~1 of \citet{2020ApJ...888....3D}. Due to the weakness of the \ion{O}{iv} lines and overlapping nature of multi-slit spectra, the lines will likely not be useful in most circumstances. However, an isolated flare kernel such as discussed in this article may yield useful spectra. In addition, the higher spatial resolution of MUSE (about a factor 10 improvement over EIS) may lead to a relative enhancement of transition region lines compared to coronal lines in the spectrum. Evidence has been found for magnetic field lines converging at low heights in flare loops \citep{2008A&A...489L..57K}. If the flare kernel corresponds to the cross-section of such a loop, and the MUSE spatial resolution is sufficient to resolve this cross-section, as would be expected from the results of \citet{2012ApJ...750L...7X} who found sizes of 0.6\as\ to 2.0\as, then transition region lines formed deeper in the atmosphere compared to coronal lines would then be expected to be relatively stronger compared to the EIS data where the kernel is likely unresolved. In the current EIS spectrum \ion{Fe}{xv} \lam284.16 (the dominant line in the MUSE 284~\AA\ channel) is a factor 15 stronger than \ion{O}{iv} \lam279.93.

\begin{acknowledgements}
    The author thanks Dr.\ Helen Mason for valuable comments on the manuscript, and Dr.\ Joel Allred for help with the F-CHROMA models. The author acknowledges funding from the GSFC Internal  Scientist Funding Model competitive work package program, the \hinode\ project, and the NASA Heliophysics Data Resource Library. \hinode\ is a Japanese mission developed and launched by ISAS/JAXA, with NAOJ as domestic partner and NASA and STFC (UK) as international partners. It is operated by these agencies in co-operation with ESA and NSC (Norway).
\end{acknowledgements}

\facilities{Hinode(EIS), SDO(AIA)}

\bibliography{ms}{}

\begin{thebibliography}{}
\expandafter\ifx\csname natexlab\endcsname\relax\def\natexlab#1{#1}\fi
\providecommand{\url}[1]{\href{#1}{#1}}
\providecommand{\dodoi}[1]{doi:~\href{http://doi.org/#1}{\nolinkurl{#1}}}
\providecommand{\doeprint}[1]{\href{http://ascl.net/#1}{\nolinkurl{http://ascl.net/#1}}}
\providecommand{\doarXiv}[1]{\href{https://arxiv.org/abs/#1}{\nolinkurl{https://arxiv.org/abs/#1}}}

\bibitem[{{Allred} {et~al.}(2015){Allred}, {Kowalski}, \&
  {Carlsson}}]{2015ApJ...809..104A}
{Allred}, J.~C., {Kowalski}, A.~F., \& {Carlsson}, M. 2015, \apj, 809, 104,
  \dodoi{10.1088/0004-637X/809/1/104}

\bibitem[{{Benz}(2017)}]{2017LRSP...14....2B}
{Benz}, A.~O. 2017, Living Reviews in Solar Physics, 14, 2,
  \dodoi{10.1007/s41116-016-0004-3}

\bibitem[{{Brosius} {et~al.}(2016){Brosius}, {Daw}, \&
  {Inglis}}]{2016ApJ...830..101B}
{Brosius}, J.~W., {Daw}, A.~N., \& {Inglis}, A.~R. 2016, \apj, 830, 101,
  \dodoi{10.3847/0004-637X/830/2/101}

\bibitem[{{Carlsson} {et~al.}(2023){Carlsson}, {Fletcher}, {Allred}, {Heinzel},
  {Ka{\v{s}}parov{\'a}}, {Kowalski}, {Mathioudakis}, {Reid}, \&
  {Sim{\~o}es}}]{2023A&A...673A.150C}
{Carlsson}, M., {Fletcher}, L., {Allred}, J., {et~al.} 2023, \aap, 673, A150,
  \dodoi{10.1051/0004-6361/202346087}

\bibitem[{{Chintzoglou} {et~al.}(2015){Chintzoglou}, {Patsourakos}, \&
  {Vourlidas}}]{2015ApJ...809...34C}
{Chintzoglou}, G., {Patsourakos}, S., \& {Vourlidas}, A. 2015, \apj, 809, 34,
  \dodoi{10.1088/0004-637X/809/1/34}

\bibitem[{{Corr{\'e}g{\'e}} \& {Hibbert}(2004)}]{2004ADNDT..86...19C}
{Corr{\'e}g{\'e}}, G., \& {Hibbert}, A. 2004, Atomic Data and Nuclear Data
  Tables, 86, 19, \dodoi{10.1016/j.adt.2003.11.002}

\bibitem[{{Culhane} {et~al.}(2007){Culhane}, {Harra}, {James}, {Al-Janabi},
  {Bradley}, {Chaudry}, {Rees}, {Tandy}, {Thomas}, {Whillock}, {Winter},
  {Doschek}, {Korendyke}, {Brown}, {Myers}, {Mariska}, {Seely}, {Lang}, {Kent},
  {Shaughnessy}, {Young}, {Simnett}, {Castelli}, {Mahmoud}, {Mapson-Menard},
  {Probyn}, {Thomas}, {Davila}, {Dere}, {Windt}, {Shea}, {Hagood}, {Moye},
  {Hara}, {Watanabe}, {Matsuzaki}, {Kosugi}, {Hansteen}, \&
  {Wikstol}}]{2007SoPh..243...19C}
{Culhane}, J.~L., {Harra}, L.~K., {James}, A.~M., {et~al.} 2007, \solphys, 243,
  19, \dodoi{10.1007/s01007-007-0293-1}

\bibitem[{{De Pontieu} {et~al.}(2020){De Pontieu}, {Mart{\'\i}nez-Sykora},
  {Testa}, {Winebarger}, {Daw}, {Hansteen}, {Cheung}, \&
  {Antolin}}]{2020ApJ...888....3D}
{De Pontieu}, B., {Mart{\'\i}nez-Sykora}, J., {Testa}, P., {et~al.} 2020, \apj,
  888, 3, \dodoi{10.3847/1538-4357/ab5b03}

\bibitem[{{De Pontieu} {et~al.}(2014){De Pontieu}, {Title}, {Lemen}, {Kushner},
  {Akin}, {Allard}, {Berger}, {Boerner}, {Cheung}, {Chou}, {Drake}, {Duncan},
  {Freeland}, {Heyman}, {Hoffman}, {Hurlburt}, {Lindgren}, {Mathur}, {Rehse},
  {Sabolish}, {Seguin}, {Schrijver}, {Tarbell}, {W{\"u}lser}, {Wolfson},
  {Yanari}, {Mudge}, {Nguyen-Phuc}, {Timmons}, {van Bezooijen}, {Weingrod},
  {Brookner}, {Butcher}, {Dougherty}, {Eder}, {Knagenhjelm}, {Larsen},
  {Mansir}, {Phan}, {Boyle}, {Cheimets}, {DeLuca}, {Golub}, {Gates}, {Hertz},
  {McKillop}, {Park}, {Perry}, {Podgorski}, {Reeves}, {Saar}, {Testa}, {Tian},
  {Weber}, {Dunn}, {Eccles}, {Jaeggli}, {Kankelborg}, {Mashburn}, {Pust},
  {Springer}, {Carvalho}, {Kleint}, {Marmie}, {Mazmanian}, {Pereira}, {Sawyer},
  {Strong}, {Worden}, {Carlsson}, {Hansteen}, {Leenaarts}, {Wiesmann},
  {Aloise}, {Chu}, {Bush}, {Scherrer}, {Brekke}, {Martinez-Sykora}, {Lites},
  {McIntosh}, {Uitenbroek}, {Okamoto}, {Gummin}, {Auker}, {Jerram}, {Pool}, \&
  {Waltham}}]{2014SoPh..289.2733D}
{De Pontieu}, B., {Title}, A.~M., {Lemen}, J.~R., {et~al.} 2014, \solphys, 289,
  2733, \dodoi{10.1007/s11207-014-0485-y}

\bibitem[{{Del Zanna}(2009)}]{2009A&A...508..501D}
{Del Zanna}, G. 2009, \aap, 508, 501, \dodoi{10.1051/0004-6361/200913082}

\bibitem[{{Del Zanna}(2012)}]{2012A&A...546A..97D}
---. 2012, \aap, 546, A97, \dodoi{10.1051/0004-6361/201219923}

\bibitem[{{Del Zanna}(2013)}]{2013A&A...555A..47D}
---. 2013, \aap, 555, A47, \dodoi{10.1051/0004-6361/201220810}

\bibitem[{{Del Zanna} {et~al.}(2021){Del Zanna}, {Dere}, {Young}, \&
  {Landi}}]{2021ApJ...909...38D}
{Del Zanna}, G., {Dere}, K.~P., {Young}, P.~R., \& {Landi}, E. 2021, \apj, 909,
  38, \dodoi{10.3847/1538-4357/abd8ce}

\bibitem[{{Dere} {et~al.}(2023){Dere}, {Del Zanna}, {Young}, \&
  {Landi}}]{2023ApJS..268...52D}
{Dere}, K.~P., {Del Zanna}, G., {Young}, P.~R., \& {Landi}, E. 2023, \apjs,
  268, 52, \dodoi{10.3847/1538-4365/acec79}

\bibitem[{{Dhakal} {et~al.}(2020){Dhakal}, {Zhang}, {Vemareddy}, \&
  {Karna}}]{2020ApJ...901...40D}
{Dhakal}, S.~K., {Zhang}, J., {Vemareddy}, P., \& {Karna}, N. 2020, \apj, 901,
  40, \dodoi{10.3847/1538-4357/abacbc}

\bibitem[{{Doschek} {et~al.}(2013){Doschek}, {Warren}, \&
  {Young}}]{2013ApJ...767...55D}
{Doschek}, G.~A., {Warren}, H.~P., \& {Young}, P.~R. 2013, \apj, 767, 55,
  \dodoi{10.1088/0004-637X/767/1/55}

\bibitem[{{Dud{\'\i}k} {et~al.}(2014){Dud{\'\i}k}, {Del Zanna},
  {Dzif{\v{c}}{\'a}kov{\'a}}, {Mason}, \& {Golub}}]{2014ApJ...780L..12D}
{Dud{\'\i}k}, J., {Del Zanna}, G., {Dzif{\v{c}}{\'a}kov{\'a}}, E., {Mason},
  H.~E., \& {Golub}, L. 2014, \apjl, 780, L12,
  \dodoi{10.1088/2041-8205/780/1/L12}

\bibitem[{{Dufresne} {et~al.}(2020){Dufresne}, {Del Zanna}, \&
  {Badnell}}]{2020MNRAS.497.1443D}
{Dufresne}, R.~P., {Del Zanna}, G., \& {Badnell}, N.~R. 2020, \mnras, 497,
  1443, \dodoi{10.1093/mnras/staa2005}

\bibitem[{{Edl\'en}(1934)}]{edlen34}
{Edl\'en}, B. 1934, Nova Acta Reg.\ Soc.\ Sci.\ Uppsala, 9, 1

\bibitem[{{Feldman} {et~al.}(1977){Feldman}, {Doschek}, \&
  {Rosenberg}}]{1977ApJ...215..652F}
{Feldman}, U., {Doschek}, G.~A., \& {Rosenberg}, F.~D. 1977, \apj, 215, 652,
  \dodoi{10.1086/155399}

\bibitem[{{Feuchtgruber} {et~al.}(1997){Feuchtgruber}, {Lutz}, {Beintema},
  {Valentijn}, {Bauer}, {Boxhoorn}, {De Graauw}, {Haser}, {Haerendel}, {Heras},
  {Katterloher}, {Kester}, {Lahuis}, {Leech}, {Morris}, {Roelfsema}, {Salama},
  {Schaeidt}, {Spoon}, {Vandenbussche}, \& {Wieprecht}}]{1997ApJ...487..962F}
{Feuchtgruber}, H., {Lutz}, D., {Beintema}, D.~A., {et~al.} 1997, \apj, 487,
  962, \dodoi{10.1086/304649}

\bibitem[{{Freeland} \& {Handy}(1998)}]{1998SoPh..182..497F}
{Freeland}, S.~L., \& {Handy}, B.~N. 1998, \solphys, 182, 497,
  \dodoi{10.1023/A:1005038224881}

\bibitem[{{Freeland} \& {Handy}(2012)}]{2012ascl.soft08013F}
---. 2012, {SolarSoft: Programming and data analysis environment for solar
  physics}, Astrophysics Source Code Library, record ascl:1208.013.
\newblock \doeprint{1208.013}

\bibitem[{{Harrison} {et~al.}(1995){Harrison}, {Sawyer}, {Carter}, {Cruise},
  {Cutler}, {Fludra}, {Hayes}, {Kent}, {Lang}, {Parker}, {Payne}, {Pike},
  {Peskett}, {Richards}, {Gulhane}, {Norman}, {Breeveld}, {Breeveld}, {Al
  Janabi}, {Mccalden}, {Parkinson}, {Self}, {Thomas}, {Poland}, {Thomas},
  {Thompson}, {Kjeldseth-Moe}, {Brekke}, {Karud}, {Maltby}, {Aschenbach},
  {Br{\"a}uninger}, {K{\"u}hne}, {Hollandt}, {Siegmund}, {Huber}, {Gabriel},
  {Mason}, \& {Bromage}}]{1995SoPh..162..233H}
{Harrison}, R.~A., {Sawyer}, E.~C., {Carter}, M.~K., {et~al.} 1995, \solphys,
  162, 233, \dodoi{10.1007/BF00733431}

\bibitem[{{Hayes} \& {Shine}(1987)}]{1987ApJ...312..943H}
{Hayes}, M., \& {Shine}, R.~A. 1987, \apj, 312, 943, \dodoi{10.1086/164939}

\bibitem[{{Kamio} {et~al.}(2010){Kamio}, {Hara}, {Watanabe}, {Fredvik}, \&
  {Hansteen}}]{2010SoPh..266..209K}
{Kamio}, S., {Hara}, H., {Watanabe}, T., {Fredvik}, T., \& {Hansteen}, V.~H.
  2010, \solphys, 266, 209, \dodoi{10.1007/s11207-010-9603-7}

\bibitem[{{Keenan} {et~al.}(2002){Keenan}, {Ahmed}, {Brage}, {Doyle}, {Espey},
  {Exter}, {Hibbert}, {Keenan}, {Madjarska}, {Mathioudakis}, \&
  {Pollacco}}]{2002MNRAS.337..901K}
{Keenan}, F.~P., {Ahmed}, S., {Brage}, T., {et~al.} 2002, \mnras, 337, 901,
  \dodoi{10.1046/j.1365-8711.2002.05988.x}

\bibitem[{{Ko} {et~al.}(2009){Ko}, {Doschek}, {Warren}, \&
  {Young}}]{2009ApJ...697.1956K}
{Ko}, Y.-K., {Doschek}, G.~A., {Warren}, H.~P., \& {Young}, P.~R. 2009, \apj,
  697, 1956, \dodoi{10.1088/0004-637X/697/2/1956}

\bibitem[{{Kontar} {et~al.}(2008){Kontar}, {Hannah}, \&
  {MacKinnon}}]{2008A&A...489L..57K}
{Kontar}, E.~P., {Hannah}, I.~G., \& {MacKinnon}, A.~L. 2008, \aap, 489, L57,
  \dodoi{10.1051/0004-6361:200810719}

\bibitem[{{Landi} {et~al.}(2010){Landi}, {Raymond}, {Miralles}, \&
  {Hara}}]{2010ApJ...711...75L}
{Landi}, E., {Raymond}, J.~C., {Miralles}, M.~P., \& {Hara}, H. 2010, \apj,
  711, 75, \dodoi{10.1088/0004-637X/711/1/75}

\bibitem[{{Landi} \& {Young}(2009)}]{2009ApJ...706....1L}
{Landi}, E., \& {Young}, P.~R. 2009, \apj, 706, 1,
  \dodoi{10.1088/0004-637X/706/1/1}

\bibitem[{{Lemen} {et~al.}(2012){Lemen}, {Title}, {Akin}, {Boerner}, {Chou},
  {Drake}, {Duncan}, {Edwards}, {Friedlaender}, {Heyman}, {Hurlburt}, {Katz},
  {Kushner}, {Levay}, {Lindgren}, {Mathur}, {McFeaters}, {Mitchell}, {Rehse},
  {Schrijver}, {Springer}, {Stern}, {Tarbell}, {Wuelser}, {Wolfson}, {Yanari},
  {Bookbinder}, {Cheimets}, {Caldwell}, {Deluca}, {Gates}, {Golub}, {Park},
  {Podgorski}, {Bush}, {Scherrer}, {Gummin}, {Smith}, {Auker}, {Jerram},
  {Pool}, {Soufli}, {Windt}, {Beardsley}, {Clapp}, {Lang}, \&
  {Waltham}}]{2012SoPh..275...17L}
{Lemen}, J.~R., {Title}, A.~M., {Akin}, D.~J., {et~al.} 2012, \solphys, 275,
  17, \dodoi{10.1007/s11207-011-9776-8}

\bibitem[{{Liang} {et~al.}(2012){Liang}, {Badnell}, \&
  {Zhao}}]{2012A&A...547A..87L}
{Liang}, G.~Y., {Badnell}, N.~R., \& {Zhao}, G. 2012, \aap, 547, A87,
  \dodoi{10.1051/0004-6361/201220277}

\bibitem[{{Liu} {et~al.}(2014){Liu}, {Richardson}, {Wang}, \&
  {Luhmann}}]{2014ApJ...788L..28L}
{Liu}, Y.~D., {Richardson}, J.~D., {Wang}, C., \& {Luhmann}, J.~G. 2014, \apjl,
  788, L28, \dodoi{10.1088/2041-8205/788/2/L28}

\bibitem[{{Mariska}(2013)}]{2013SoPh..282..629M}
{Mariska}, J.~T. 2013, \solphys, 282, 629, \dodoi{10.1007/s11207-012-0200-9}

\bibitem[{{Milligan}(2011)}]{2011ApJ...740...70M}
{Milligan}, R.~O. 2011, \apj, 740, 70, \dodoi{10.1088/0004-637X/740/2/70}

\bibitem[{{Muglach} {et~al.}(2010){Muglach}, {Landi}, \&
  {Doschek}}]{2010ApJ...708..550M}
{Muglach}, K., {Landi}, E., \& {Doschek}, G.~A. 2010, \apj, 708, 550,
  \dodoi{10.1088/0004-637X/708/1/550}

\bibitem[{{Polito} {et~al.}(2016){Polito}, {Del Zanna}, {Dud{\'\i}k}, {Mason},
  {Giunta}, \& {Reeves}}]{2016A&A...594A..64P}
{Polito}, V., {Del Zanna}, G., {Dud{\'\i}k}, J., {et~al.} 2016, \aap, 594, A64,
  \dodoi{10.1051/0004-6361/201628965}

\bibitem[{{Sandlin} {et~al.}(1986){Sandlin}, {Bartoe}, {Brueckner}, {Tousey},
  \& {Vanhoosier}}]{1986ApJS...61..801S}
{Sandlin}, G.~D., {Bartoe}, J. D.~F., {Brueckner}, G.~E., {Tousey}, R., \&
  {Vanhoosier}, M.~E. 1986, \apjs, 61, 801, \dodoi{10.1086/191131}

\bibitem[{{Tripathi} {et~al.}(2009){Tripathi}, {Mason}, {Dwivedi}, {del Zanna},
  \& {Young}}]{2009ApJ...694.1256T}
{Tripathi}, D., {Mason}, H.~E., {Dwivedi}, B.~N., {del Zanna}, G., \& {Young},
  P.~R. 2009, \apj, 694, 1256, \dodoi{10.1088/0004-637X/694/2/1256}

\bibitem[{{Warren} {et~al.}(2008){Warren}, {Feldman}, \&
  {Brown}}]{2008ApJ...685.1277W}
{Warren}, H.~P., {Feldman}, U., \& {Brown}, C.~M. 2008, \apj, 685, 1277,
  \dodoi{10.1086/591075}

\bibitem[{{Warren} {et~al.}(2014){Warren}, {Ugarte-Urra}, \&
  {Landi}}]{2014ApJS..213...11W}
{Warren}, H.~P., {Ugarte-Urra}, I., \& {Landi}, E. 2014, \apjs, 213, 11,
  \dodoi{10.1088/0067-0049/213/1/11}

\bibitem[{{Widing} \& {Cook}(1987)}]{1987ApJ...320..913W}
{Widing}, K.~G., \& {Cook}, J.~W. 1987, \apj, 320, 913, \dodoi{10.1086/165609}

\bibitem[{{Xu} {et~al.}(2012){Xu}, {Cao}, {Jing}, \&
  {Wang}}]{2012ApJ...750L...7X}
{Xu}, Y., {Cao}, W., {Jing}, J., \& {Wang}, H. 2012, \apjl, 750, L7,
  \dodoi{10.1088/2041-8205/750/1/L7}

\bibitem[{{Young}(2021)}]{2021FrASS...8...50Y}
{Young}, P.~R. 2021, Frontiers in Astronomy and Space Sciences, 8, 50,
  \dodoi{10.3389/fspas.2021.662790}

\bibitem[{Young(2022{\natexlab{a}})}]{young_2022_7255646}
Young, P.~R. 2022{\natexlab{a}}, Calibrating EIS data: the EIS\_PREP routine,
  3.8,  Zenodo, \dodoi{10.5281/zenodo.7255646}

\bibitem[{Young(2022{\natexlab{b}})}]{young_2022_6339584}
---. 2022{\natexlab{b}}, {EIS\_AUTO\_FIT and SPEC\_GAUSS\_EIS: Gaussian fitting
  routines for the Hinode/EIS mission}, 3.1,  Zenodo,
  \dodoi{10.5281/zenodo.6339584}

\bibitem[{Young(2023{\natexlab{a}})}]{peter_r_young_2023_8097457}
---. 2023{\natexlab{a}}, {CHIANTI Technical Report No. 3: Computing a synthetic
  spectrum with CHIANTI}, 1.5,  Zenodo, \dodoi{10.5281/zenodo.8097457}

\bibitem[{Young(2023{\natexlab{b}})}]{young_2023_10371938}
---. 2023{\natexlab{b}}, {Co-aligning the EIS flare data from 9 March 2012 with
  AIA images}, 1.0,  Zenodo, \dodoi{10.5281/zenodo.10371938}

\bibitem[{{Young}(2023)}]{2023ApJ...958...40Y}
{Young}, P.~R. 2023, \apj, 958, 40, \dodoi{10.3847/1538-4357/ad0548}

\bibitem[{{Young} {et~al.}(2016){Young}, {Dere}, {Landi}, {Del Zanna}, \&
  {Mason}}]{2016JPhB...49g4009Y}
{Young}, P.~R., {Dere}, K.~P., {Landi}, E., {Del Zanna}, G., \& {Mason}, H.~E.
  2016, Journal of Physics B Atomic Molecular Physics, 49, 074009,
  \dodoi{10.1088/0953-4075/49/7/074009}

\bibitem[{{Young} {et~al.}(2013){Young}, {Doschek}, {Warren}, \&
  {Hara}}]{2013ApJ...766..127Y}
{Young}, P.~R., {Doschek}, G.~A., {Warren}, H.~P., \& {Hara}, H. 2013, \apj,
  766, 127, \dodoi{10.1088/0004-637X/766/2/127}

\bibitem[{{Young} {et~al.}(2018){Young}, {Keenan}, {Milligan}, \&
  {Peter}}]{2018ApJ...857....5Y}
{Young}, P.~R., {Keenan}, F.~P., {Milligan}, R.~O., \& {Peter}, H. 2018, \apj,
  857, 5, \dodoi{10.3847/1538-4357/aab556}

\bibitem[{{Young} \& {Landi}(2009)}]{2009ApJ...707..173Y}
{Young}, P.~R., \& {Landi}, E. 2009, \apj, 707, 173,
  \dodoi{10.1088/0004-637X/707/1/173}

\bibitem[{{Young} \& {Mason}(1997)}]{1997SoPh..175..523Y}
{Young}, P.~R., \& {Mason}, H.~E. 1997, \solphys, 175, 523,
  \dodoi{10.1023/A:1004936106427}

\bibitem[{{Young} {et~al.}(2012){Young}, {O'Dwyer}, \&
  {Mason}}]{2012ApJ...744...14Y}
{Young}, P.~R., {O'Dwyer}, B., \& {Mason}, H.~E. 2012, \apj, 744, 14,
  \dodoi{10.1088/0004-637X/744/1/14}

\bibitem[{{Young} {et~al.}(2021{\natexlab{a}}){Young}, {Ryabtsev}, \&
  {Landi}}]{2021ApJ...908..104Y}
{Young}, P.~R., {Ryabtsev}, A.~N., \& {Landi}, E. 2021{\natexlab{a}}, \apj,
  908, 104, \dodoi{10.3847/1538-4357/abd39b}

\bibitem[{{Young} \& {Ugarte-Urra}(2022)}]{2022SoPh..297...87Y}
{Young}, P.~R., \& {Ugarte-Urra}, I. 2022, \solphys, 297, 87,
  \dodoi{10.1007/s11207-022-02014-4}

\bibitem[{{Young} {et~al.}(2021{\natexlab{b}}){Young}, {Viall}, {Kirk},
  {Mason}, \& {Chitta}}]{2021SoPh..296..181Y}
{Young}, P.~R., {Viall}, N.~M., {Kirk}, M.~S., {Mason}, E.~I., \& {Chitta},
  L.~P. 2021{\natexlab{b}}, \solphys, 296, 181,
  \dodoi{10.1007/s11207-021-01929-8}

\bibitem[{{Young} {et~al.}(2007){Young}, {Del Zanna}, {Mason}, {Dere}, {Landi},
  {Landini}, {Doschek}, {Brown}, {Culhane}, {Harra}, {Watanabe}, \&
  {Hara}}]{2007PASJ...59S.857Y}
{Young}, P.~R., {Del Zanna}, G., {Mason}, H.~E., {et~al.} 2007, \pasj, 59,
  S857, \dodoi{10.1093/pasj/59.sp3.S857}

\end{thebibliography}
\bibliographystyle{aasjournal}

\appendix

\section{Radiometric calibration uncertainties}\label{app.calib}

An important source of uncertainties for line ratio measurements lies in the EIS radiometric calibration. \citet{2013SoPh..282..629M} found that the instrument sensitivity had declined after launch, and modeled an exponential decay that was the same at all wavelengths. Further work by GDZ13 and WUL14 identified a wavelength dependence to the sensitivity decay, with the LW channel showing a stronger decay than the SW channel. The authors also derived modifications to the original calibration within the two channels. Although the results of GDZ13 and WUL14 are broadly similar, there are important differences and the \href{https://solarb.mssl.ucl.ac.uk/JSPWiki/Wiki.jsp?page=EISAnalysisGuide}{EIS data analysis guide} does not recommend one over the other.

The key \ion{O}{iv} lines used in the present work are found between 268 and 280~\AA\ and Table~\ref{tbl.calib} shows the effect of the GDZ13 and WUL14 calibrations on ratio values of 1.0 in the pre-launch calibration. The WUL14 values are very similar, but the GDZ14 values are around 15\%\ lower. Also shown is the \ion{O}{v} \lam192.90/\lam248.46 ratio, which is used in Section~\ref{sect.comp}. Since the lines are in different channels, the updated ratios are significantly different from the pre-launch calibration values. The GDZ13 and WUL14 values are also very different, with the GDZ13 ratio 36\,\%\ lower.

\begin{deluxetable}{llcccc}[t]
\tablecaption{Ratio comparison for different radiometric calibrations.\label{tbl.calib}}
\tablehead{
    \colhead{Ion} &
    \colhead{Ratio} &
    \colhead{Theory} & 
    \colhead{Pre-launch} & 
    \colhead{GDZ13} &
    \colhead{WUL14}
}
\startdata
\ion{O}{v} & \lam192.90/\lam248.46 & \nodata & 1.0 & 0.57 & 0.37 \\
\ion{O}{iv} & \lam207.16/\lam279.93 & \nodata & 1.0 & 0.50 & 0.40 \\
\ion{O}{iv} & \lam268.02/\lam279.93 & \nodata & 1.0 & 0.87 & 1.00 \\
 & \lam272.13/\lam279.93 & \nodata & 1.0 & 0.84 & 1.04 \\
\noalign{\smallskip}
\ion{Si}{x} & \lam277.26/\lam271.99 & 0.77 & 0.83 & 0.95 & 0.80 \\
\enddata
\end{deluxetable}

Also shown in Table~\ref{tbl.calib} is a \ion{Si}{x} branching ratio, \lam277.26/\lam271.99, which takes a constant value of 0.77 in all plasma conditions (data from CHIANTI). Since the two lines are close in wavelength to the \ion{O}{iv} lines then they can be used to estimate the accuracy of the radiometric calibration options in this region. The two lines were measured in a quiet Sun off-limb dataset from 2011 August 14, beginning at 05:33~UT, which was the nearest such dataset to the flare observation time. The spectrum and line fit parameters are available on Zenodo at DOI:10.5281/zenodo.6726426.  Table~\ref{tbl.calib} gives the measured ratios using the three calibration options. The WUL14 calibration gives the best agreement, with the GDZ13 value more than 20\%\ higher than the expected value. For this reason, the WUL14 calibration is used in the present work. However, an additional 5\%\ uncertainty is added to the measured line intensities to represent the uncertainties due to the radiometric calibration in this wavelength region.

\end{document}